\documentclass{article}
\usepackage{jcappub}
\usepackage{lmodern}
\usepackage{graphicx}
\usepackage{amsfonts}
\usepackage{amsmath}
\hypersetup{colorlinks = true, urlcolor = blue, linkcolor = blue, frenchlinks = true, pdfborder = {0 0 0}}

\begin{document}

\title{On Geometrization of Spinors in a Complex Spacetime}

\author{Santanu Das}
\affiliation{Imperial College London,\\Exhibition Rd, South Kensington, London SW7 2BX}
\emailAdd{santanu.das@imperial.ac.uk}
\abstract{
While general relativity provides a complete geometric theory of gravity, it fails to explain the other three forces of nature, i.e., electromagnetism and weak and strong interactions. We require the quantum field theory (QFT) to explain them. Therefore, in this article, we try to geometrize the spinor fields. We define a parametric coordinate system in the tangent space of a null manifold and show that these parametric coordinates behave as spinors. By introducing a complex internal hyperspace on a tangent space of a null manifold, we show that we can get eight sets of such parametric coordinate systems that can behave as eight spinor fields. These spinor fields contain two triplets that can rotate among themselves under SU(3). Seven of these spinor fields also couple with a U(1) field with different strengths. We also show that while these spinors can be assigned a tensor weight $1/2$ or $-1/2$, in a $L^p$ space where the coordinates, instead of adding up in quadrature, add up in $p$th power, we can get a parametric space that contains similar spinors of tensor weight $1/p$.  
} 
\maketitle

\section{Introduction}

The general theory of relativity (GR) is the best-known theory of gravity. It is a geometric approach to gravity. However, this simple geometric approach is only limited to gravity; and does not explain the other three forces of nature. Instead, to explain these forces, we need the quantum field theory. In particle physics, we break the particles into two main categories. The elementary Bosons consist of photons, gluons, and W and Z bosons as the force carrier and the recently discovered scalar particle Higgs Boson. On the other hand, the elementary Fermions consist of quarks and leptons. There exist three generations of these particles which are carbon copies of one another, except for their mass. Each generation has two leptons, such as $e$ and $\nu_e$ for the first generation. There are also two types of quarks for each generation, e.g., $u$ and $d$ for the first generations. The quarks can carry $3$ color charges, namely red, green, and blue. So independently for the first generation, there are total $6$ quarks, $3$ $u$s or $u_r$, $u_g$ and $u_b$  and $3$ $d$s, namely, $d_r$, $d_b$, $d_g$. Combining with leptons, there are a total of $3+3+2=8$ types of Fermions. Each of these Fermions comes in two different forms based on their helicity -- right-handed and left-handed (considering that right-handed neutrinos exist). Each particle also has its anti-particles, giving total $2\times 2 \times 8=32$ forms of particles for each generation of Fermions. In the QFT, the Fermions do not arise naturally. Instead, we need to add the Fermions into the equations separately. 

In general relativity a space-time is a Lorentzian manifold $(\mathcal{M},g)$, i.e. a smooth ( infinitely differentiable ) manifold $\mathcal{M}$ with a globally defined tensor field $g:T\mathcal{M}\times T\mathcal{M}\rightarrow \mathbb{R}$. $T\mathcal{M}$ is the tangent space on $\mathcal{M}$.
Suppose $X \in T\mathcal{M}$, then according to the general theory of relativity, the quantity $g_{ab}X^aX^b$, $\forall a,b=0,..,3$ remains constant under the change of the coordinate system. This is a quadratic relation  
 and it consists of two sets of linear equations. Earlier authors in \cite{Fock-1929, Veblen2-1933,Veblen4-1933} showed that we could parameterize the set of linear equations on a quadratic curve using two complex numbers. A 2-dimensional coordinate system defined using these complex numbers behaves as massless spinors. This approach is also used by Penrose and others in the twister formalism~\cite{penrose1967twistor,penrose1973twistor}. 

In this paper, instead of going to the twister approach,  we define a parametric coordinate system to represent spinors. We then introduce an internal complex $S^3$ space on the tangent space of a null manifold; and define eight independent sets of such parametric-space coordinate systems. Among these eight sets of parametric space coordinates, we get two triplets that can rotate within themselves under SU($3$) transformation. Seven sets of these parametric space coordinate systems can couple with a U($1$) field with different coupling strengths -- one of them couples with a unit strength, three others from one triplet can couple with $\frac{1}{3}$ of the strength and rest three from the other triplet couple with the same U($1$) field with a strength of $\frac{2}{3}$. 
This, in a way, relates the three color charges of quantum chromodynamics (QCD) to the three spatial dimensions in our coordinate system. 

We also show that we can assign a tensor weight $\frac{1}{2}$ to these parametric space coordinates. One interesting aspect is to note that while our universe is governed by a space where the coordinates add up quadratically, there is no apparent mathematical reason why we cannot mathematically consider an $L^p$ space and a tensor field  $g:\left(T\mathcal{M}\right)^p\rightarrow \mathbb{R}$, such that $g_{a_1a_2 ... a_p}X^{a_1}X^{a_2}... X^{a_p}$ remains unchanged under any coordinate transformation. In this article, we show that in such a space, we can define a similar complex parametric space coordinate system that can have a tensor weight of $\frac{1}{p}$.

This article is organized as follows. In Sec.~\ref{sectionTangentSpace}, we briefly review the works of Fock, Veblen and others; and define a parametric space coordinate system over the tangent space of a 4-dimensional null hyper-surface~\cite{Fock-1929, Veblen2-1933,Veblen4-1933,History_of_UFT,History_of_UFT_2}. The following two sections describe the co-variant derivative, parallel transport, additional degrees of freedom, etc. In Sec.~\ref{sec6}, we define $8$ different sets of parametric space coordinate systems and discuss their inter transformations through gauge rotation. Finally, we have a discussion and conclusion. The appendix shows how a similar analysis can be done on an $L^p$ space that can lead to a spin $\frac{1}{p}$ system. We also provide appendices to help the reader better understand the rotation in the complex internal space and visualize SU($3$) rotation.

\section{\label{sectionTangentSpace}Mathematical setup for tangent space and parametric space coordinates}

\subsection{\label{SLE_section}The system of linear equations}

Any quadratic can be expressed as a system of linear equations. The most basic example is a circle,  
\begin{equation}
    X^2 + Y^2 = K^2\,.
\end{equation}

\noindent $K\in \mathbb{R}$ is the radius of the circle. As $(K+X)(K-X) = Y^2$, this can be written as a pair of straight lines as 
\begin{eqnarray}
\qquad X = K + \lambda Y\,,  \qquad\qquad
\qquad X = K - \frac{1}{\lambda}Y\,.
\end{eqnarray}
\noindent Here $\lambda$ is a parameter. Given any $\lambda\in \mathbb{R} - (0,\infty,-\infty)$, we get one line from each of the set, which meet at a point,  $\mathcal{P}(\lambda)$, the locus of which gives the circle. 

Provided $K$ is a variable, lets say $K = T$, the circle becomes a cone, i.e. $X^2 + Y^2 = T^2$. In this case, each of the systems of linear equations represents a plane through the origin. These planes cut each other in a straight line, and these systems of straight lines together create the cone. Therefore, any point $\lambda$ represents a unique straight line on the cone. 

Let us consider a hyperboloid by adding some constant term, $K$, with the cone

\begin{equation}
 X^{2}+Y^{2}-T^{2}=K^2   \,.
\end{equation}

\noindent Here, the lines are difficult to visualize in the first place. As usual, the intersection of a plane with a quadratic yields a  curve. However, that is not the case here; the hyperboloid contains straight lines. For instance, the line $(K,\,0,\,0)+\lambda(0,\,1,\,1)$ is on the hyperboloid, and it can be tested by substituting the point $(K,\,\lambda,\,\lambda)$ on the hyperboloid. 
To check it mathematically, we can rewrite the equation as
\begin{equation}
 (X+T)(X-T)=(K+Y)(K-Y)   \,.
\end{equation}
For a given number, say $\mu$, the planes $X+T=\mu (K+Y)$ and $\mu (X-T)=(K-Y)$ cut each other on the hyperboloid,  giving one set of lines. We can consider another parameter, $\lambda$, to get the other set of lines. The planes $(X+T)=\lambda (K-Y)$ and $\lambda (X-T)=(K+Y)$ produces another sets of lines on the hyperboloid. No two lines of the same group intersect with each other, and any two lines of the opposite groups intersect at one and only one point. Therefore, any set $(\mu, \lambda)$ represents  a particular point $\mathcal{P}_2(\lambda, \mu)$, on the hyperboliod. 

If we allow to vary $K$, let us say $K = Z$, the set of straight lines becomes a set of planes. and the point $\mathcal{P}_2(\lambda, \mu)$ now becomes a straight line. 
All these lines meet each other only at the origin, and they span the entire space. For any line, we have a unique $(\lambda, \mu)$, and the location on that straight line can be given by the value of $Z$. 

Here we should note that if we take a hyperboloid of the form $T^2 = X^2 + Y^2 + Z^2$, then the system of planes becomes imaginary; however, the lines still remain real, as $( T, X, Y, Z )$  are real. The entire space is spanned by a series of straight lines, each of which can be parameterized by two parameters $(\lambda, \mu)$. Note that these $\lambda$ and $\mu$ may be real or complex numbers.

\subsection{Defining tangent space coordinate system}
Consider a 4-dimension null spacetime manifold, $\mathcal{M}$, represented by the coordinate system $(x^0,x^1,x^2,x^3)$ and a line element is given by $g_{\mu\nu}dx^{\mu}dx^{\nu} = 0$ on the manifold, where $\mu,\nu \in (0,\ldots 3)$. 
At each point, $\mathcal{P}$, of this manifold there is a tangent space $T\mathcal{M}$,  whose coordinates are $\frac{dx^\mu}{d\lambda}$. In  tangent space, this coordinates system  follows the quadratic  $g_{\mu\nu}\frac{dx^\mu}{d\lambda}\frac{dx^\nu}{d\lambda} = 0$. $\lambda$ is an affine parameter on the null hyperspace. The tangent space is flat. Hence, we can define a Minkowski coordinate system on the tangent space, given by $(X^0,X^1,X^2,X^3)$. At any point $\mathcal{P}$, on the manifold, if we transform the coordinate system $\frac{dx}{d\lambda}\rightarrow X$, then these new coordinates should follow the quadratic (conic section)

\begin{equation}
\label{conic_section}
\left(X^{0}\right)^{2}-\left(X^{1}\right)^{2}-\left(X^{2}\right)^{2}-\left(X^{3}\right)^{2}=0\,,
\end{equation}

\noindent which can also be written as $\eta_{ab}X^{a}X^{b}=0$, where $\eta_{ab}$ is the Minkowskian metric, and $a,b\in (0,\ldots 3)$ are the indices to represent the coordinate system on the tangent space. 

The transformation from the $\frac{dx^\mu}{d\lambda}$ coordinate system to $X^a$ coordinate can be written as
\begin{equation}
X^{a} = \Lambda_{\mu}^{a} \frac{dx^{\mu}}{d\lambda} \qquad\textrm{and hence}\qquad \eta_{ab}\Lambda_{\mu}^{a}\Lambda_{\nu}^{b} = g_{\mu\nu}.     
\label{a_mu_relation}
\end{equation}

\noindent For the shake of mathematical benefit, we can define
\begin{equation}
\Lambda^{\mu a} = \Lambda^{a \mu} = g^{\mu\nu}\Lambda_{\nu}^{a}
\qquad\textrm{and}\qquad \Lambda_{a \mu}=\eta_{ab}\Lambda^{b}_{\mu}\,.
\end{equation}
\noindent This also gives 

\begin{equation}
 \Lambda^{\mu a}\Lambda_{\nu a}=g^{\mu\lambda}\Lambda_{\lambda}^{a}\eta_{ab}\Lambda^{b}_{\nu} = 
g^{\mu\lambda}g_{\lambda\nu} = \delta_{\nu}^{\mu}   
\end{equation}
\noindent and

\begin{equation}
\Lambda_{\nu}^{\mu}=\Lambda_{a}^{\mu}\Lambda_{\nu}^{a}=\delta_{\nu}^{\mu}
\qquad\textrm{and} \qquad
\Lambda_{b}^{\mu}\Lambda_{\mu}^{a}=\Lambda_{b}^{a}=\delta_{b}^{a}\,.
\end{equation}

\noindent Here, we are free to assume any coordinate transformation in the $x^\mu$ reference frame. It does not change the tangent space. For any transformation in $x^\mu$, each $a$ component of $\Lambda_{a}^{\mu}$, i.e. $\Lambda_{0}^{\mu},\Lambda_{1}^{\mu},\Lambda_{2}^{\mu},\Lambda_{3}^{\mu}$  transforms as contravariant vectors and each $a$ component of $\Lambda^{a}_{\mu}$ transforms as a co-variant vector. 
On the other hand, the tangent space coordinates can be subjected to any Lorentz transformation. 

\begin{equation}
\bar{X}^{a}=L_{b}^{a}X^{b} \qquad\textrm{and hence}\qquad \bar{\Lambda}^a_\mu = L^a_b\Lambda^b_\mu
\end{equation}

\subsection{Parameterizing the tangent space \label{tangent_space}}

For parameterizing the tangent space coordinates, we follow the discussion from Sec.~\ref{SLE_section}. 
We start with Eq.~\ref{conic_section}, which can be written in the determinant form as 

\begin{equation}
\left|\begin{array}{cc}
X^{0}+X^{3} & X^{1}+iX^{2}\\
X^{1}-iX^{2} & X^{0}-X^{3}
\end{array}\right|= 0\,. 
\label{eq:Determinant}
\end{equation}

\noindent The determinant of a $2\times 2$ matrix is zero if and only if its two rows are equal up to some scaling factor. Therefore, if the determinant has to vanish, then we must have variables like $\psi^{1}$,
$\psi^{2}$, $\psi^{3}$, $\psi^{4}$, such that 

\begin{equation}
\begin{array}{ccc}
X^{0}+X^{3}=\psi^{1}\psi^{3}\,, &  &  
X^{1}+iX^{2}=\psi^{1}\psi^{4}\,, \\
X^{1}-iX^{2}=\psi^{2}\psi^{3}\,, &  &  
X^{0}-X^{3}=\psi^{2}\psi^{4}\,. 
\end{array}
\label{parametricform38}
\end{equation}

\noindent $\psi^{1}$, $\psi^{2}$, $\psi^{3}$, $\psi^{4}$ are complex variables. 

In Sec.~\ref{SLE_section}, we discuss two systems of planes, described by two sets of linear equations, which are parameterized by two parameters, $\lambda$, and $\mu$. For our parameterization in Eq.~\ref{parametricform38}, the first sets of planes can be obtained by the intersection of the linear equations $\left(X^{1}+iX^{2}\right) = \left(\psi^{4} / \psi^{3}\right)\left(X^{0}+X^{3}\right)$
and $\left(X^{0}-X^{3}\right)=\left(\psi^{4} / \psi^{3}\right) \left( X^{1}-iX^{2}\right)$. 
This plane remains unaltered if $\psi^{3}$ and $\psi^{4}$ are multiplied by the same factor. There is one and only one such plane for each value of this ratio  $\psi^{3}/\psi^{4}$, and if the ratio changes, the plane changes. Similarly, the other system of linear equations on the cone, i.e. 
 $\left(X^{1}+iX^{2}\right) = \left(\psi^{1} / \psi^{2}\right)\left(X^{0}-X^{3}\right)$
and $\left(X^{0}+X^{3}\right)=\left(\psi^{1} / \psi^{2}\right) \left( X^{1}+iX^{2}\right)$
is represented by the value of $\psi^{1}/\psi^{2}$. Each plane from the system cuts each plane of the other in a straight line. These systems of straight lines span the entire space as given by Eq.~\ref{eq:Determinant}. Therefore,  $\psi^{1}$, $\psi^{2}$ and $\psi^{3}$, $\psi^{4}$ give a parametric representation of the tangent space light cone.

As  $\psi^{1}$, $\psi^{2}$, $\psi^{3}$, and $\psi^{4}$ are complex quantities, we have total $8$ degrees of freedom, i.e., $5$ extra degrees of freedom from the original equation, which has $4$ variables and one constraint. Imposing the reality condition, i.e. $(X^0 + X^3)$ and $(X^0 - X^3)$ are real,  we should get 

\begin{equation}
\psi^{*3}=k\psi^{1}
\qquad\textrm{ and }\qquad
\psi^{*4}=k\psi^{2}\,,
\end{equation}

\noindent where we are using the $*$ to indicate complex conjugates and $k$ is a real number. The effect of changing the value of $k$ is shifting the point $X^a$ along the same line on the light cone. Hence, no generality is lost by assuming $k=1$. The points $X^a$s on the light cone are given by 

\begin{eqnarray}
X^{1}=\frac{\psi^{1}\psi^{*2}+\psi^{2}\psi^{*1}}{2}\,, \qquad\qquad
X^{2}=\frac{\psi^{1}\psi^{*2}-\psi^{2}\psi^{*1}}{2i}\,, \nonumber \\
X^{3}=\frac{\psi^{1}\psi^{*1}-\psi^{2}\psi^{*2}}{2}\,,\qquad\qquad
X^{0}=\frac{\psi^{1}\psi^{*1}+\psi^{2}\psi^{*2}}{2}\,.
\label{XtoPhiRelationFull}
\end{eqnarray}

This reduces the degrees of freedom to $4$. We still have one extra degree of freedom, which we will discuss in detail in Sec.~\ref{phase} of this article. Each point on the tangent space $T\mathcal{M}$, can be parameterized by the quantity $\psi^{A}$, where $A\in (1,2)$. 
In this article, we refer $\psi^A$ as a parametric space coordinate. The indices are referred to as a spin index, and we use the roman capital letters, $A, B, ...$ to represent the spin indices of $\psi^A$. 

The transformation relation between $\psi^A$ and $X^a$ coordinate system can be written as 

\begin{equation}
X^a = \mathcal{G}^a_{AB}\psi^A\psi^{*B}\,,
\label{XtoPhi}
\end{equation}

\noindent where $\mathcal{G}^a_{AB}$ are the matrices 

\begin{eqnarray}
\mathcal{G}^0_{AB} = \frac{1}{2}\begin{pmatrix}
1 & 0 \\
0 & 1 
\end{pmatrix} \,,\quad
\mathcal{G}^1_{AB} = \frac{1}{2}\begin{pmatrix}
0 & 1 \\
1 & 0 
\end{pmatrix}\,,\quad
\mathcal{G}^2_{AB} = \frac{1}{2i}\begin{pmatrix}
0 & 1 \\
-1 & 0 
\end{pmatrix}\,,\quad
\mathcal{G}^3_{AB} = \frac{1}{2}\begin{pmatrix}
1 & 0 \\
0 & -1 
\end{pmatrix}\,.
\label{Gi_to_AB}
\end{eqnarray}

\noindent We can also write a reverse transformation as 
\begin{equation}
\psi^A\psi^{*B} = \mathcal{G}_a^{AB} X^a \,.
\label{PhiToX}
\end{equation}
\noindent Here $\mathcal{G}_a^{AB}$ are the matrix inverse of the matrices in Eq.~\ref{Gi_to_AB}. This leads us to the relations 

\begin{equation}
    \mathcal{G}_a^{AB}\mathcal{G}^b_{AB} = \delta^b_a   \qquad\qquad
    \mathcal{G}_a^{AB}\mathcal{G}^a_{CD} = \delta^A_C\delta^B_D
\end{equation}

\noindent We must note that $\mathcal{G}_a^{AB}$s are not symmetric. The first index is used for the non-starred coordinates, and the second is for the starred coordinates. 

\subsection{\label{transformationlaws32}Transformation laws for parametric space  coordinates}

If we subject $\psi^{1}$ and $\psi^{2}$
to a linear transformation, then we get

\begin{equation}
\bar{\psi^{A}}=T_{B}^{A}\psi^{B}\,.\label{eq:Transform}
\end{equation}

\noindent It can be easily verified that $\bar{\psi}^A$ also satisfy the quadratic relation given by Eq.~(\ref{eq:Determinant}). However, the equation gets multiplied by the square of the determinant

\begin{equation}
T=\left|T_{B}^{A}\right|\,.
\end{equation}

\noindent We may consider the Eq.(\ref{eq:Transform}) as a transformation of the parametric coordinates $\psi^{A}$ into new coordinates $\bar{\psi}^{C}$, i.e. as a change of reference system. Under such linear transformation, $X^a$s also undergo a linear transformation which can be verified by simple algebraic manipulation. 

For providing $\psi^A$s with a tensor-like structure, we define their covariant counterparts as $\psi_A$. Subjected to the transformation of Eq.~\ref{eq:Transform}, these covariant components also go through a covariant transformation as 

\begin{equation}
\bar{\psi}_{A}=t_{A}^{B}\psi_{B}\,\,\,\,\,\,\,\,\,\,\,\,\,\,(A,\,B\,=1,2)
\qquad\text{where,}\qquad
t_{B}^{A}T_{C}^{B}=\delta_{C}^{A}\,.
\end{equation}
\noindent 
 
Let us define a quantity $\epsilon_{AB}$, that can lower the index of $\psi^A$. The inverse of this matrix $\epsilon^{AB}$ can be used to raise the indices. As we use $\epsilon_{AB}$ to lower the indices, in the  barred and unbarred reference frame, it should satisfy

\begin{equation}
    \epsilon_{AB}\psi^A\psi^B = \bar{\epsilon}_{CD}\bar{\psi}^C\bar{\psi}^D =\bar{\epsilon}_{CD}T^C_A T^D_B \psi^A \psi^B\,, \qquad\qquad \text{giving,} \qquad\qquad 
    \epsilon_{AB} =\bar{\epsilon}_{CD}T^C_A T^D_B \,.
    \label{eabtransform318}
\end{equation}

\noindent Provided we take $\epsilon^{AB}$, $\epsilon_{AB}$ to be the Levi-Civita symbols, i.e.

\begin{eqnarray}
\epsilon^{11}=\epsilon^{22}=0\,,
\qquad\quad
\epsilon^{12}=-\epsilon^{21}=1\,, 
\qquad\quad
\epsilon_{11}=\epsilon_{22}=0\,,
\qquad\quad
\epsilon_{12}=-\epsilon_{21}=1 \,,
\end{eqnarray}

\noindent and we demand that $\epsilon_{AB}$ are invariant under such linear transform, then the left-hand side of Eq.~\ref{eabtransform318} get multiplied with $T$, i.e., we get $\epsilon_{AB} ={\epsilon}_{CD}T^{-1}T^C_A T^D_B$. If we take $t = |t^A_B|$ then we can write the equations in terms of $t$ as

\begin{equation}
\epsilon_{CD}=\frac{1}{t}\epsilon_{AB}t_{C}^{A}t_{D}^{B}\,,
\qquad\qquad
\epsilon^{CD}=t\epsilon^{AB}T_{A}^{C}T_{B}^{D}\,.
\label{Relation_Of_la_visa_Beta}
\end{equation}

\noindent Therefore, $\epsilon_{AB}$ and $\epsilon^{AB}$ behave as a covariant tensor density of weight $-1$ and contravariant tensor density of weight $1$ respectively~\footnote{ A tensor density or relative tensor is a generalization of tensor. A tensor density transforms as a tensor when coordinate system transformation except that it is additionally multiplied by a power $w$ of the Jacobian determinant of the coordinate transition function.

Considering an arbitrary transformation from a general coordinate system to another, a relative tensor of weight $w$ is defined by the following tensor transformation:
$$
\bar{A}^{i j \ldots k}_{l m \ldots n}=\left|\frac{\partial x}{\partial \bar{x}}\right|^{w} \frac{\partial \bar{x}^{i}}{\partial x^{a}} \frac{\partial \bar{x}^{j}}{\partial x^{b}} \cdots \frac{\partial \bar{x}^{k}}{\partial x^{c}} \frac{\partial x^{d}}{\partial \bar{x}^{l}} \frac{\partial x^{e}}{\partial \bar{x}^{m}} \cdots \frac{\partial x^{f}}{\partial \bar{x}^{n}} A^{a b \ldots c}{ }_{d e \ldots f}
$$

This is useful in calculating area or volume. We can get an example from the cross-product.

$\vec{u} \times \vec{v}=\left[\begin{array}{ll}u_{1} & u_{2}\end{array}\right]\left[\begin{array}{cc}0 & 1 \\ -1 & 0\end{array}\right]\left[\begin{array}{l}v_{1} \\ v_{2}\end{array}\right]=u_{1} v_{2}-u_{2} v_{1}$

This is invariant because it represents the area of a triangle, which should not change due to the change in the coordinate system. Now let under the change of coordinate system, $\bar{u}=A^{-1} u$ and $\bar{v}=A^{-1} v$. Therefore, under such coordinate transformation  $\left(A^{-1}\right)^{\top}\left[\begin{array}{cc}0 & 1 \\ -1 & 0\end{array}\right] A^{-1}$
 the original expression but multiplied by $\operatorname{det} A^{-1}$. This could be thought of as a two-index tensor transformation, but instead, it is computationally easier to think of the tensor densities.
}.
Raising the indices by $\epsilon^{AB}$ increases the weight
by $+1$, and lowering indices by $\epsilon_{AB}$ decreases it by $-1$. If we assign
$\psi^{A}$ to be of weight $+\frac{1}{2}$ then $\psi_{A}$ must
be of weight $-\frac{1}{2}$.  Assuming that $\psi^{A}$ are of weight $+\frac{1}{2}$, linear transformation of the $\psi^A$-coordinate can be re-written as 

\begin{equation}
\bar{\psi}^{A}=t^{\frac{1}{2}}T_{B}^{A}\psi^{B}\,.
\label{PsiTransform}
\end{equation}

\noindent This transformation has a unit determinant and does not multiply the light cone (given by Eq.~\ref{conic_section}) by $T^2$. In this article, we refer to such transformation as the spin-space coordinate transformation or simply the spin transformation. 

Raising and lowering indices work as follows.

\begin{equation}
\epsilon_{BA}\psi^{B}=\psi_{A}\,,
\qquad\qquad
\epsilon^{AB}\psi_{B}=\psi^{A}\,,
\end{equation}

\noindent which implies 

\begin{equation}
\psi_{1}=-\psi^{2}\,,
\qquad\qquad
\psi_{2}=\psi^{1}\,.
\end{equation}

\noindent Here one should note that $\epsilon$ is not a symmetric metric. We use the first index to raise the indices and the second to lower them. This raising and lowering of the indices and the choice of $\epsilon_{AB}$ are purely mathematical constructs for simplifying calculations. Note that, we can also define such parametric space coordinates of spin $\frac{1}{p}$ in the tangent space of an $L^p$ space. Check Appendix~\ref{Lp_space} for a detailed discussion.

Suppose at any point $\mathcal{P}$ on a null manifold $\mathcal{M}$, and the line element is given by $g_{\mu\nu}dx^\mu dx^\nu = 0$ and $\lambda$ is an affine parameter on the null manifold. The  4-velocity of a particle on the manifold is given by $\frac{dx^\mu}{d\lambda}$. 
If the tangent space coordinates at $\mathcal{P}$ are given by $X^a$, then these coordinates essentially represent the four-velocity of the particle at $\mathcal{P}$ and not the real spacetime coordinates. 
We relate $X^a$'s with the 4-velocities as $X^a = \Lambda^a_\mu \frac{dx^\mu}{d\lambda}$, such that $X^a$s satisfy Eq.~\ref{conic_section} and the curvature of the spacetime gets absorbed in $\Lambda^a_\mu$. Therefore, these tangent space coordinates are free from any gravitation effect. We define the parametric space coordinate system, i.e., $\psi^A$, on the tangent space, and hence they are also free from any spacetime curvature at that point. 

As $\psi^A$ are complex numbers, we can write them as  $\psi^1=|\psi^1|\exp(i\theta^1)$ and $\psi^2 = |\psi^2|\exp(i\theta^2)$, where $|\psi^A|$ are the amplitudes and $\theta^A$ are the arguments of the complex quantities. This gives
\begin{eqnarray}
X^{1} &= |\psi^1||\psi^2|\cos(\theta^1 - \theta^2)\,, \quad
X^{2} &= |\psi^1||\psi^2|\sin(\theta^1 - \theta^2)\,, \quad \nonumber\\
X^{3} &= \frac{1}{2}\left(|\psi^1|^2 - |\psi^2|^2 \right)\,, \quad
X^{0} &= \frac{1}{2}\left(|\psi^1|^2 + |\psi^2|^2 \right)\,. 
\end{eqnarray}

\noindent Given $X^a$, we can uniquely determine $|\psi^1|$, $|\psi^2|$ and $(\theta^1 - \theta^2)$. However, we have one degree of freedom unfixed. Without fixing one of $\theta^1$ or $\theta^2$, $\psi^A$'s can not be determined uniquely. 
These additional degrees of freedom play an essential role in gauge choice and we discuss these additional degrees of freedom in Sec.~\ref{phase}.

\section{Covariant differentiation of the $\psi$ system}

In accordance with our calculatioin in the previous section, $\psi^A$ behaves as a tensor density of weight $\frac{1}{2}$ and $\psi_A$ as a weight $-\frac{1}{2}$. Hence, co-variant derivative of $\psi^A$ can be written as\footnote{
If $A^i$ is a relative tensor of weight $w$, then the covariant derivative of the $A^i$ can be written as 
\begin{equation}
    A_{ ; q}^{i}=\partial_{q} A^{i} +\Gamma_{a q}^{i} A^{a}-w A^{i} \Gamma_{a q}^{a}\,,
\end{equation}
i.e., we need to add the negative term with a normal covariant derivative if it was a tensor. Here $\Gamma_{a q}^{i}$ are the connection parameters, which are also known as the Crystoffel symbol for Romanian space} 

\begin{equation}
\psi_{;\alpha}^{A} = 
\frac{\partial \psi^{A}}{\partial x^{\alpha}}+\Upsilon_{B \alpha}^{A} \psi^{B}-\frac{1}{2} \Upsilon_{B \alpha}^{B} \psi^{A}\end{equation}

\noindent where $\Upsilon_{B \alpha}^{A}$ is a connection parameter. The $\frac{1}{2}$ comes in the second term because of the weight of $\psi^A$. Under coordinate transformation, $\psi_{;\alpha}^{A}$ must transform as a vector with respect to $\alpha$ and transform as spinor ( transform as Eq.~\ref{PsiTransform}) with respect to gauge and the spin transform ( by gauge transform we mean the change of the extra free parameter of the system, which is discussed in Sec.~\ref{phase} ).
 
\begin{equation}
\left(\frac{\partial \bar{\psi}^{A}}{\partial \bar{x}^{\alpha}}+\bar{\Upsilon}_{C \alpha}^{A} \bar{\psi}^{C}-\frac{1}{2} \bar{\Upsilon}_{C \alpha}^{C} \bar{\psi}^{A}\right)
=t^{1 / 2}  T_{B}^{A} \frac{\partial x^{\beta}}{\partial \bar{x}^{\alpha}}
\left(\frac{\partial \psi^{B}}{\partial x^{\beta}}+\Upsilon_{D \beta}^{B} \psi^{D}-\frac{1}{2} \Upsilon_{D \beta}^{D} \psi^{A}\right)
\label{particle_transformation}
\end{equation}

\noindent Let us consider $\Gamma^{A}_{B\alpha} = \Upsilon_{B \alpha}^{A} - \frac{1}{2}\Upsilon_{C \alpha}^{C}\delta^A_B$. Therefore, in terms of this new variable, the covariant derivative of $\Psi^A$ can be written as 

\begin{equation}
\psi_{;\alpha}^{A} = 
\frac{\partial \psi^{A}}{\partial x^{\alpha}}+\Gamma_{B \alpha}^{A} \psi^{B}\,.
\end{equation}

\noindent If we take the complex conjugate of $\psi^{A}$, we should get similar equations for the complex conjugates. Therefore, we can write 

\begin{eqnarray}
    (\psi^A\psi^{*B})_{;\alpha} &=& 
    \psi^A\left( \frac{\partial \psi^{*B}}{\partial x^{\alpha}}+\Gamma_{C \alpha}^{*B} \psi^{*C} \right) + \left( \frac{\partial \psi^{A}}{\partial x^{\alpha}}+\Gamma_{C \alpha}^{A} \psi^{C} \right) \psi^{*B} \nonumber \\
    &=&  \frac{ \partial\left(\psi^A\psi^{*B}\right)}{\partial x^{\alpha}}  + \Gamma_{C \alpha}^{*B} \psi^A\psi^{*C} + \Gamma_{C \alpha}^{A} \psi^{C} \psi^{*B}\,.
\end{eqnarray}

\noindent We can multiply this equation with $\mathcal{G}^a_{AB}$ and use Eq.\ref{PhiToX} and Eq.~\ref{XtoPhi} to get 

\begin{eqnarray}
X^a_{;\alpha} &=& \frac{ \partial X^a}{\partial x^{\alpha}}  + \mathcal{G}^a_{AB}\Gamma_{C \alpha}^{*B} \psi^A\psi^{*C} + \mathcal{G}^a_{BC}\Gamma_{A \alpha}^{B} \psi^{A} \psi^{*C} \noindent \nonumber\\
&=& \frac{ \partial X^a}{\partial x^{\alpha}}  + \left(\mathcal{G}^a_{AB}\Gamma_{C \alpha}^{*B}  + \mathcal{G}^a_{BC}\Gamma_{A \alpha}^{B}\right) \mathcal{G}_b^{AC} X^{b} 
= \frac{ \partial X^a}{\partial x^{\alpha}}  + \Gamma^a_{b\alpha} X^b\,,
\end{eqnarray}

\noindent Here, $X^a$ are the tangent space coordinates. $\Gamma^a_{b\alpha}$ are connection parameters for the covariant derivative of $X^a$ coordinates. Therefore, we can relate the connection parameter of the parametric space with that of the tangent space coordinate system as

\begin{equation}
    \Gamma^a_{b\alpha} = \mathcal{G}^a_{AB}\Gamma_{C \alpha}^{*B}\mathcal{G}_b^{AC}  + \mathcal{G}^a_{BC}\Gamma_{A \alpha}^{B}\mathcal{G}_b^{AC}\,.
    \label{Connection_AB_2_ab}
\end{equation}

We know that the tangent space coordinates $X^a$ are related to the coordinate of the spacetime manifold, i.e., $dx^\mu$  as $\frac{dx^\mu}{d\lambda} = \Lambda^\mu_a X^a$. Therefore, we can relate the Christoffel symbols on the manifold with the connections, $\Gamma^a_{b\alpha} $ as 

\begin{equation}\Gamma_{b \alpha}^{a}=\left(\Gamma_{\nu \alpha}^{\mu} \Lambda_{\mu}^{a}-\frac{\partial \Lambda_{\nu}^{a}}{\partial x^{\alpha}}\right) \Lambda_{b}^{\nu}\,.
\label{Gamma-alpha-a}
\end{equation}

\noindent Using simple algebraic manipulations, this relation can also be inverted, which is  given by 

\begin{equation}
\Gamma_{\nu \alpha}^{\mu} = 
\left(\Gamma_{b \alpha}^{a} \Lambda^{b}_{\nu} +\frac{\partial \Lambda_{\nu}^{a}}{\partial x^{\alpha}}\right)\Lambda^{\mu}_{a} \,. \end{equation}

\noindent Inverting the relation given in Eq.~\ref{Connection_AB_2_ab} is tricky. We can multiply both sides with $\mathcal{G}_a^{PQ}$ for inverting it.

\begin{eqnarray}
    \Gamma^a_{b\alpha}\mathcal{G}_a^{PQ} &=& \mathcal{G}_a^{PQ}\mathcal{G}^a_{AB}\Gamma_{C \alpha}^{*B}\mathcal{G}_b^{AC}  + \mathcal{G}_a^{PQ}\mathcal{G}^a_{BC}\Gamma_{A \alpha}^{B}\mathcal{G}_b^{AC} \nonumber \\
    &=& \delta^{P}_{A}\delta^{Q}_{B}\Gamma_{C \alpha}^{*B}\mathcal{G}_b^{AC}  + \delta^{P}_{B}\delta^{Q}_{C}\Gamma_{A \alpha}^{B}\mathcal{G}_b^{AC} 
    = \Gamma_{C \alpha}^{*Q}\mathcal{G}_b^{PC}  + \Gamma_{A \alpha}^{P}\mathcal{G}_b^{AQ} \,.
\end{eqnarray}

\noindent Multiplying it with another $\mathcal{G}^b_{RS}$ we get

\begin{eqnarray}
\mathcal{G}^b_{RS}\Gamma^a_{b\alpha}\mathcal{G}_a^{PQ} &=&\Gamma_{C \alpha}^{*Q}\mathcal{G}_b^{PC}\mathcal{G}^b_{RS}  + \Gamma_{A \alpha}^{P}\mathcal{G}_b^{AQ}\mathcal{G}^b_{RS} \nonumber \\
&=&\Gamma_{C \alpha}^{*Q}\delta^{P}_{R}\delta^{C}_{S}  + \Gamma_{A \alpha}^{P}\delta^{A}_{R}\delta^{Q}_{S} = \Gamma_{S \alpha}^{*Q}\delta^{P}_{R}  + \Gamma_{R \alpha}^{P}\delta^{Q}_{S}\,.
\label{Eq_515}
\end{eqnarray}

\noindent Considering $Q=S$ we get
\begin{equation}
\frac{1}{2}\mathcal{G}^b_{R(Q)}\Gamma^a_{b\alpha}\mathcal{G}_a^{P(Q)} 
 = \Gamma_{(Q) \alpha}^{*(Q)}\delta^{P}_{R}  + \Gamma_{R \alpha}^{P}\,.
 \label{Gamma-a-A}
\end{equation}

\noindent 
Note that we replace $S$ with $Q$, with no summation over $Q$ in the $\Gamma$. We put a bracket around $Q$ to the point that out. We can assume $\Gamma_{(Q) \alpha}^{(Q)} = C_\alpha$, where $C_\alpha$ is a complex vector field.  
This gives $\Gamma_{R \alpha}^{P} = \frac{1}{2}\mathcal{G}^b_{RQ}\Gamma^a_{b\alpha}\mathcal{G}_a^{PQ} - C^{*}_{\alpha}\delta^{P}_{R}$. However, we also need to satisfy the relation Eq.~\ref{Eq_515}, which requires $C^*_\alpha = -C_\alpha$. Therefore, the complex vector field has to be imaginary, giving us $C_\alpha = -iA_\alpha$, where $A_\alpha$ is a real vector field. Therefore, we can write.

\begin{equation}
    \Gamma_{R \alpha}^{P} = \frac{1}{2}\mathcal{G}^b_{RQ}\Gamma^a_{b\alpha}\mathcal{G}_a^{PQ} + iA_{\alpha}\delta^{P}_{R}\,.
    \label{Eq517}
\end{equation}

Knowing the Christoffel symbols on a manifold, we can find $\Gamma^a_{b\alpha}$ and calculate $\Gamma^P_{R\alpha}$. However, one interesting thing is that the covariant derivative of $\psi^A$ is connected with an additional vector field that does not come from the spacetime curvature. 

\subsection{Rule of transformations for the connection parameters}

For calculating the rule of transformation for the connection parameter $\Gamma^P_{R\alpha}$ and $\Upsilon^{P}_{R\alpha}$, we first differentiate both sides of the first relation of 
Eq.~\ref{Relation_Of_la_visa_Beta} with respect to $x^\alpha$
\begin{equation}
    0=\epsilon_{AB}\frac{\partial}{\partial x^\alpha} (\frac{1}{t})t_{C}^{A}t_{D}^{B}+ \epsilon_{AB}\frac{1}{t}\frac{\partial}{\partial x^\alpha}(t_{C}^{A})t_{D}^{B} + \epsilon_{AB}\frac{1}{t}t_{C}^{A}\frac{\partial}{\partial x^\alpha}(t_{D}^{B})\,.
\end{equation}

\noindent Multiplying this with the second relation of the same equation set, we get 

\begin{eqnarray}
    0&=&\epsilon_{AB}\frac{\partial}{\partial x^\alpha} (\frac{1}{t})t_{C}^{A}t_{D}^{B}t\epsilon^{RS}T_{R}^{C}T_{S}^{D}+ \epsilon_{AB}\frac{1}{t}\frac{\partial}{\partial x^\alpha}(t_{C}^{A})t_{D}^{B}t\epsilon^{RS}T_{R}^{C}T_{S}^{D} + \epsilon_{AB}\frac{1}{t}t_{C}^{A}\frac{\partial}{\partial x^\alpha}(t_{D}^{B})t\epsilon^{RS}T_{R}^{C}T_{S}^{D} \nonumber \\
    &=& \epsilon_{AB}\frac{\partial}{\partial x^\alpha} (\frac{1}{t})t\epsilon^{RS} \delta^A_R \delta^B_S+ \epsilon_{AB}\frac{\partial}{\partial x^\alpha}(t_{C}^{A})\epsilon^{RS}T_{R}^{C} \delta^B_S+ \epsilon_{AB}\frac{\partial}{\partial x^\alpha}(t_{D}^{B})\epsilon^{RS}T_{S}^{D}\delta^A_R \nonumber \\
    &=& 2t \frac{\partial}{\partial x^\alpha} (\frac{1}{t})+4T^C_A\frac{\partial t^A_C}{\partial x^\alpha}\,.
\end{eqnarray}

\noindent As $\psi^B$ transform as $t^{\frac{1}{2}}T^A_B\psi^B$, we get

\begin{eqnarray}
\frac{\partial \bar{\psi}^B}{\partial x^\alpha} = \frac{\partial \left(t^\frac{1}{2}T^B_A\psi^A\right)}{\partial x^\alpha} = 
t^\frac{1}{2}T^B_A \frac{\partial \psi^A}{\partial x^\alpha} +
t^\frac{1}{2}\psi^A\frac{\partial T^B_A}{\partial x^\alpha} + 
\frac{1}{2}T^B_A\psi^At^{\frac{1}{2}}\frac{\partial \ln t}{\partial x^\alpha}\,.
\end{eqnarray}

\noindent However, $\bar{\psi}^A_{;\beta}$ is the covariant derivative. Therefore, under the spin transformation, it should transform as $\bar{\psi}^B_{;\alpha}=t^{\frac{1}{2}}T^B_A\psi^A_{;\alpha}$. Using some simple algebraic manipulations, we can show that under coordinate and spin transformation, the connection parameters, $\Gamma_{B \beta}^{A}$ transform as 

\begin{equation}
\label{Gamma_matrix_spin_transform}
\bar{\Gamma}_{B \beta}^{A}=\left[\left(\Gamma_{D \alpha}^{C} t_{B}^{D}+\frac{\partial t_{B}^{C}}{\partial x^{\alpha}}\right) T_{C}^{A}-\frac{1}{2} \frac{\partial \ln t}{\partial x^{\alpha}} \delta_{B}^{A}\right]\frac{\partial x^\alpha}{\partial \bar{x}^\beta}\,.
\end{equation}

\noindent As $\Gamma^A_{B\alpha} = \Upsilon^A_{B\alpha} - \frac{1}{2}\Upsilon^C_{C\alpha}\delta^A_B$, to satisfy Eq.~\ref{Gamma_matrix_spin_transform},  the variable $\Upsilon^A_{B\alpha}$ must transform as   

\begin{equation}
\bar{\Upsilon}_{D \beta}^{C}=\left(\Upsilon_{B \alpha}^{A} t_{D}^{B}+\frac{\partial t_{D}^{A}}{\partial x^{\alpha}}\right) T_{A}^{C} \frac{\partial x^{\alpha}}{\partial \bar{x}^{\beta}}\,.
\label{UpsilonTransform}
\end{equation}

\subsection{Calculating the parallel transport equations}

The above section discusses the covariant derivative of the $\psi^A$ coordinate system. We can use these covariant derivatives to calculate the parallel transport equation for the $\psi^A$ system, i.e., if we move the particle freely from one point on the spacetime to another, how will the parametric coordinate system change. 

If we assume $V^\mu = \frac{dx^\mu}{d\lambda}$, then the parallel transport equation for the vector $V^\mu$ is
given by 

\begin{equation}
\frac{dV^\mu}{d\lambda} + \Gamma^\mu_{\nu\rho}V^\rho\frac{dx^\nu}{d\lambda} = 0
\end{equation}

\noindent According to Eq.~\ref{a_mu_relation}, the quantities $X^a$ and  $V^\mu$ are related as $V^\mu = \Lambda^\mu_a X^a$.  Thus, the parallel transport equations for $X^a$'s look like

\begin{eqnarray}
\frac{d(\Lambda^\mu_a X^a) }{d\lambda} + \Gamma^\mu_{\nu\rho}\Lambda^\mu_a X^a\frac{dx^\nu}{d\lambda} = 0 \\
\Rightarrow \Lambda^\mu_a \frac{d X^a }{d\lambda} + X^a \frac{d\Lambda^\mu_a}{d\lambda} + \Gamma^\mu_{\nu\rho}\Lambda^\mu_a X^a\frac{dx^\nu}{d\lambda} = 0 
\end{eqnarray}

\noindent Using Eq.~\ref{Gamma-alpha-a}, we get the equation for the parallel transport of $X^a$ as

\begin{equation}
\frac{dX^a}{d\lambda} + \Gamma^a_{b\nu}X^b\frac{dx^\nu}{d\lambda} = 0\,.
\end{equation}

\noindent For calculating the parallel transport equations for $\Psi^A$s, we can use the relation Eq.~\ref{XtoPhi}.  Considering  $\mathcal{G}^a_{AB}$ as constants we get

\begin{equation}
\frac{d(\psi^A\psi^{*B})}{d\lambda} + \mathcal{G}_a^{AB}\Gamma^a_{b\nu}\mathcal{G}^b_{CD}\psi^C\psi^{*D}\frac{dx^\nu}{d\lambda} = 0
\end{equation}

\noindent Using Eq.~\ref{Eq_515} we get 

\begin{eqnarray}
\psi^A\frac{d\psi^{*B}}{d\lambda} + \psi^{*B}\frac{d\psi^A}{d\lambda} + (\Gamma_{D \nu}^{*B}\delta^{A}_{C}  + \Gamma_{C \nu}^{A}\delta^{B}_{D})\psi^C\psi^{*D}\frac{dx^\nu}{d\lambda} = 0 \\
\Rightarrow \psi^A\left(\frac{d\psi^{*B}}{d\lambda} 
+ \Gamma_{D \nu}^{*B}\psi^{*D}\frac{dx^\nu}{d\lambda}\right)
+ \psi^{*B}\left(\frac{d\psi^A}{d\lambda} 
+ \Gamma_{C \nu}^{A}\psi^C\frac{dx^\nu}{d\lambda}\right) = 0
\end{eqnarray}

\noindent Here, we have two similar equations, which are just the complex conjugate to each other, and the sum is zero. To separate these, we again need to add and subtract some quantity like $iB_\mu\psi^A \psi^{*B}$ as it was done in Eq.~\ref{Eq517}, giving 

\begin{equation}
\frac{d\psi^{A}}{d\lambda} 
+ \Gamma_{B \nu}^{A}\psi^{B}\frac{dx^\nu}{d\lambda} = iB_\mu\psi^{A}\frac{dx^\nu}{d\lambda} 
\end{equation}

\noindent This gives the equation for parallel transport for the $\Psi^A$. $B_\nu$ is an arbitrary vector field. We can take $B_\mu = 0$ as it can be absorbed in the definition of $ \Gamma_{A \nu}^{A}$, which already contains an arbitrary vector field as given in Eq.~\ref{Eq517}. 

\section{Exploring the additional degrees of freedom\label{phase}}

As $\psi^{1}$ and $\psi^{2}$ are complex numbers, there are total $4$ degrees of freedom. However, as we are trying to parameterize the tangent space of a $4$ dimensional manifold and there is one constraint, we have a total of $3$ degrees of freedom. Therefore, we have introduced one additional degree of freedom in our parameterization of the tangent space. This can be seen as follows. In Eq.~\ref{XtoPhi}, if $\psi^{1}$ and $\psi^{2}$ are multiplied with a number $c$ and $\psi^{1*}$ and $\psi^{2*}$ are multiplied with a number $\frac{1}{c}$ then $X^a$ remains unchanged. This implies that if we take $cc^*=1$, or 

\begin{equation}
c = \exp(i\theta)\,, \quad\quad\quad\text{where}\quad \theta \in \mathbb{R}\,
\end{equation}

\noindent then $X^a$s remain unchanged under transformation $\psi^A \rightarrow e^{i\theta}\psi^A$. Here we must note that no arbitrary choice of $c$ can alter the tangent space coordinate system. Therefore, $c$ does not have to be constant. It can vary over spacetime, i.e., over $x^\mu$ for $\mu \in (0,3)$. Let us consider a new coordinate transformation given by  

\begin{equation}
{\theta}'\rightarrow\theta+\rho\left(x^{0},x^{1},x^{2},x^{3}\right)\,.
\end{equation}

\noindent $\psi^A$ transform as ${\psi}'^{A} \rightarrow e^{i\theta + i\rho\left(x^{0},x^{1},x^{2},x^{3}\right)}\psi^{A}$.  ${\psi}'^A$ still gives the same unique point on the light cone as that of $\psi^A$. In this article, we refer to these types of transformations in $\psi^A$ as gauge transformation and name the coordinate $\theta$ as internal coordinate. \footnote{Note that in this article, we use the primed coordinate system for the gauge transformation and the barred coordinate system for the spin transformation.} We can further generalize the transformation by writing $\rho\left(x^{0},x^{1},x^{2},x^{3}\right)$ as a path integration from point $\mathcal{P}_0 = \{x_0^\nu |\nu\in(0,\ldots, 3)\}$ to $\mathcal{P} = \{x^\nu |\nu\in(0,\ldots, 3)\}$ as

\begin{equation}
{\theta}'\rightarrow\theta+\int_{\mathcal{P}_0}^{\mathcal{P}}\lambda_{\mu}dx^{\mu}\,,
\end{equation}

\noindent where $\lambda_\mu$ is an arbitrary vector field. This gives 

\begin{equation}
{\psi}'^{A} \rightarrow \psi^{A}\exp\left(i\theta + i\int_{\mathcal{P}_0}^{\mathcal{P}}\lambda_{\mu}dx^{\mu}\right)\,.
\end{equation}

\noindent Here, we introduce a parameterization of the light cone that depends on the path of the integration without changing the point on the light cone. Suppose we take two points $\mathcal{P}_0$ and $\mathcal{P}$ on a manifold $\mathcal{M}$. The tangent space at these two points of the manifold is fixed. Suppose a particle moves from $\mathcal{P}_0$ to $\mathcal{P}$ through two different paths, and the initial and the final points on the tangent space are the same. That does not ensure that the parametric coordinate, $\psi^A$, of the particle remains the same because the path through which the particle reaches from the first to the second point is not the same. At every point on the path, there is a vector field $\lambda_\mu$, and the phase of the coordinate $\psi^A$ depends on the path integral through which it moves from one point to another. In other words, the information of the vector field $\lambda_\mu$ on the path remains encoded on the $\psi^A$ coordinate system. 

\subsection{Holomorphicity of the {$\psi^A$} coordinates}

We have shown that $\psi^A$ are the complex functions of the tangent-space coordinates $X^a$ and the internal coordinate $\theta$, i.e. $\psi^A = \psi^A(X^a,\theta)$. Given values of $\psi^A$, we can uniquely determine $(X^a,\theta)$ and vice versa. There is an interesting property of $\psi^A$. If we club the internal coordinate $\theta$ with any of the tangent space coordinates and create a complex coordinate $Z^a = X^a \exp(i2\theta)$, then $\psi^A$ is a holomorphic function of $Z^a$. 

\paragraph{Proof:}

Let us consider $\psi^1 = |\psi^1|\exp(2i\phi_1)$ and $\psi^2 = |\psi^2|\exp(2i\phi_2)$. Let $\theta = \phi_1 + \phi_2$ and $\phi = \phi_1 - \phi_2$, where in accordance with Eq.~\ref{XtoPhiRelationFull}, $\phi = \tan^{-1} \left({X^2}/{X^1}\right)$. So using $\theta$ and $\phi$ we can write $\psi^A$ coordinates as $\psi^1 = |\psi^1|\exp(i\theta + i\phi)$ and $\psi^2 = |\psi^2|\exp(i\theta - i\phi)$. 

As $X^a$ are on the Minkowski space
\begin{eqnarray}
\label{cauchy_first_part}
\frac{ G^a_{AB}\psi^A\psi^{*B}}{X^b}=\frac{X^a}{X^b} = \frac{\partial X^a}{\partial X^b}
=\frac{\partial \left(G^a_{AB}\psi^A\psi^{*B}\right)}{\partial X^b}\,.
\end{eqnarray}

\noindent Calculating  Eq.~\ref{cauchy_first_part} for $a=0$ and $a=3$ and summing them up we get

\begin{alignat}{2}
&&\frac{ \psi^1\psi^{*1}}{X^b}
&=\frac{\partial \left(\psi^1\psi^{*1}\right)}{\partial X^b}
=\frac{\psi^{*1}\partial \psi^1}{\partial X^b} + 
\frac{ \psi^1\partial\psi^{*1}}{\partial X^b}
=2\frac{\psi^{*1}\partial \psi^1}{\partial X^b} \\
&\Rightarrow &\frac{\psi^1}{X^b}
&=2\frac{\partial \psi^1}{\partial X^b} \,.
\end{alignat}

\noindent This same relation will also be valid for $\psi^2$. Putting $\psi^1 = |\psi^1|\exp(i\theta + i\phi)$ in the above equation we get
\begin{eqnarray}
\frac{|\psi^1|\left[\cos(\theta + \phi) + i\sin(\theta + \phi)\right]}{X^b}
=2\left(\frac{\partial |\psi^1|}{\partial X^b}\right)\left[\cos(\theta + \phi) + i\sin(\theta + \phi)\right] \qquad\qquad\qquad \nonumber\\
+ 2|\psi^1|\left(\frac{\partial (\theta + \phi)}{\partial X^b}\right)\left[-\sin(\theta + \phi) 
+ i\cos(\theta + \phi)\right]  \,.  
\end{eqnarray}

\noindent Separating the real and the imaginary parts we get
\begin{eqnarray}
\frac{|\psi^1|}{X^b}\cos(\theta + \phi)
=2\frac{\partial |\psi^1|}{\partial X^b}\cos(\theta + \phi) -
2|\psi^1| \frac{\partial (\theta + \phi)}{\partial X^b} \sin(\theta + \phi)\,,  
\label{Cauchy–Riemann1} \\
\frac{|\psi^1|}{X^b}\sin(\theta + \phi)
=2\frac{\partial |\psi^1|}{\partial X^b}\sin(\theta + \phi) +
2|\psi^1|\frac{\partial (\theta + \phi)}{\partial X^b}\cos(\theta + \phi) \,.
\label{Cauchy–Riemann2}
\end{eqnarray}

\noindent Now, $\phi$ is a function of $X^a$ but $\theta$ is a completely free parameter. Consider a complex variable $Z^a = X^a \exp(2i\theta)$. We can write the following derivatives as 
\begin{eqnarray}
\frac{\partial \left(|\psi^1|\cos(\theta + \phi)\right)}{\partial X^a} &=& \frac{\partial|\psi^1|}{\partial X^a}\cos(\theta + \phi) - |\psi^1| \frac{\partial (\theta + \phi)}{\partial X^a} \sin(\theta + \phi) \,,\\
\frac{\partial\left(|\psi^1|\sin(\theta + \phi)\right)}{X^a\partial \theta} &=& 
\frac{|\psi^1|\cos(\theta + \phi)}{X^a} \,,\\
\frac{\partial \left(|\psi^1|\sin(\theta + \phi)\right)}{\partial X^a} &=& \frac{\partial|\psi^1|}{\partial X^a}\sin(\theta + \phi) + |\psi^1| \frac{\partial (\theta + \phi)}{\partial X^a} \cos(\theta + \phi) \,,\\
-\frac{\partial\left(|\psi^1|\cos(\theta + \phi)\right)}{X^a\partial \theta} &=& 
\frac{|\psi^1|\sin(\theta + \phi)}{X^a}\,.
\label{CR14}
\end{eqnarray}

\noindent Using Eq.~\ref{Cauchy–Riemann1} - Eq.~\ref{CR14}, we can see that $\psi^1$ follows the Cauchy–Riemann equation, i.e. 

\begin{eqnarray}
\frac{\partial \left(|\psi^1|\cos(\theta + \phi)\right)}{\partial X^a} = \frac{\partial\left(|\psi^1|\sin(\theta + \phi)\right)}{X^a\partial \theta} \,, \qquad
\frac{\partial \left(|\psi^1|\sin(\theta + \phi)\right)}{\partial X^a} = 
-\frac{\partial\left(|\psi^1|\cos(\theta + \phi)\right)}{X^a\partial \theta}\,.\nonumber
\end{eqnarray}

\noindent Hence, $\psi^1$ is a holomorphic function of $Z^a=X^a \exp(i2\theta)$. Similar analysis with $\psi^2$ ( subtracting $a=3$ component from $a=0$ in  Eq.~\ref{cauchy_first_part} and redoing the above calculations ), we can show that $\psi^2$ is also a holomorphic function of $Z^a$. This concludes the proof. 

According to our previous discussion, $\theta$ is an additional degree of freedom that the $\psi^A$ coordinate system preserves, along with the $X^A$ coordinates. Mathematically, there is no restriction that the $\theta$ has to be a single number. Instead of considering a single $\theta$, if we consider four independent angles $\theta^0, \theta^1$, $\theta^2$, $\theta^3$ such that $\theta= \theta^0 + \theta^1 +\theta^2 +\theta^3$ and define $Z^a = X^a\exp(2i\theta^a)$ where  $a\in (0,\ldots,3)$, then $\psi^A$ becomes holomorphic functions of $Z^a$. 

Therefore, under this setup, the tangent space coordinate axis are complex coordinates, and our parameter space coordinates $\psi^A$ are the functions of the complex coordinate systems. $\psi^A$ preserve the value of the total phase $\theta$. However, it does not independently save the phases associated with each coordinate axis. The complex phases with the coordinate axis allow us to store an independent quantity along each coordinate axis. As an explainer, we can take an example of an electromagnetic field. While analyzing EM waves, we can write the electric field as a complex number simplifying different equations. The real part of the complex number gives us the actual electric field. Similarly, for the complex coordinate systems, we can consider that the amplitude of the complex coordinate $Z^a$ can be associated with some real coordinate, and the argument of the coordinate can be associated with some additional quantity when $\theta^a \ne 0$.

Here we should also note that the tangent space of the 4-dimensional null hyperspace is a 3-dimensional space. Therefore, instead of $4$ independent values of $\theta^a$, we may consider that there are only $3$ independent $\theta$s, and the fourth value is somehow related to the other three values.

\section{\label{sec6}Exploring internal coordinate system}
\subsection{Defining different differential forms}

The parametric coordinate $\psi^A$ parameterizes the tangent space coordinate system $X^a$ and the sum of the angles $\theta^a$. These angles create a 3-sphere ($S^3$) internal subspace at every point on the tangent space (assuming that there are only three independent $\theta^a$s). The parametric coordinate $\psi^A$ can not individually parameterize $\theta^a$s. However, they pose some interesting properties. As all the $\theta^a$ angles appear in the exponent, the function remains the same if we integrate or differentiate any of these parametric space coordinates, $\psi^A$, with respect to any $\theta^a$. Therefore,  we can construct differential forms $\psi^A_p e^p$, $\psi^A_{pq} e^p \wedge e^q$ and $\psi^A_{pqr} e^p \wedge e^q \wedge e^r$, which are analytically same as $\psi^A$ with added phase. Here $e_p = \hat\theta_p$ are the basis vector along the $\theta^p$ coordinates, and the basis in the covector space is $e^{p}$ where $p \in (1,2,3)$. Here wedge represents the wedge product of the vector spaces, $e^p \wedge e^p = 0$ and $e^p \wedge e^q = - e^q \wedge e^p$ for $p\ne q$. We use the indices $p$, $q$, $r, \ldots$ to represent the $\theta^p$ coordinates in the internal $S^3$ space. These indices can vary from 1 to 3, while the tangent space indices, represented by $a$, $b, \ldots$, can vary from $0$ to $3$. We also refer these indices $p$, $q$, $r, \ldots$ as gauge index.  
These differential forms can also be used as parameters for parameterizing the light cone. However, these new parameters have directional dependence in the internal $S^3$ space. 

Here we ignore the $a=0$ component because the null hyperspace is a 3D surface. One is not independent among the $4$ coordinates in the tangent space. So, without the loss of generality, we consider $3$ spacial coordinates as independent coordinates and associate a complex phase with each space coordinate axis. The time coordinate behaves differently from the spatial coordinates as it comes with a negative signature. So, we consider it as a dependent coordinate. There should be some complex phase associated with the time coordinate too. However, we consider that phase in the time axis will somehow be related to the other three phases. We can write the light cone as $(X^0)^2 - \sum_{p=1}^{3}(Z^pZ^{*p}) = 0$. We use $Z^p = X^p\exp(2i\theta^p)$ as defined in the previous section.

Let us take the differential 0-form, 1-form, 2-form, 3-form on $S^3$ as 

\begin{alignat}{2}
&\Omega_0 : \psi^A  && \qquad\, \quad \\
&\Omega_1 : \psi^A_pe^p  && \qquad\cdots \quad\forall p\in (1,2,3)\\
&\Omega_2 : \psi^A_{pq}e^p\wedge e^q   && \qquad\cdots \quad\forall p, q\in (1,2,3) \;\;\;\;\;\;\&\; p, q\, \text{in circular order}\\
&\Omega_3 : \psi^A_{pqr}e^p\wedge e^q\wedge e^r  && \qquad\cdots \quad\forall  p, q, r\in (1,2,3)\;\;\;\&\; p<q<r
\label{diffenrential_forms_series_1}
\end{alignat}

\noindent Under this construct, the $0^{th}$ form is the standard $\psi^A$, used in the previous section, and the tangent space coordinates from this are given by Eq.~\ref{XtoPhi}. The $1-$forms have three directions in the internal $S^3$ space. We can take the dot product of them in the internal space. $\psi^A_p$ are complex numbers. However, the total strength of $\psi^A_p$, i.e. $\sum_{p=1}^3\psi^{*A}_p\psi^{A}_p$ for $\forall A\in(1,2)$ component remains constant under any type of rotation of $S^3$. For any particular $A\in (1,2)$, the magnitude of each $p$ component i.e. $\sqrt {\psi^{*A}_p\psi^A_p}$ for $p\in(1,2,3)$, rotates as SO($3$) and rotation of complex $\psi^A_p$ follows the rotation group  SU($3$) ~(check Appendix~\ref{appendixB} and  Appendix~\ref{appendixC} for further discussion). For different $p$ components, we can get the contribution to the tangent space coordinate system as $X^a=\mathcal{G}^a_{AB}\psi^{A}_p\psi^{*B}_p$ for each $p\in (1,2,3)$. The total contribution from the all three components is $X^a=\sum_{p=1}^3 \mathcal{G}^a_{AB}\psi^{A}_p\psi^{*B}_p$. Under any SU($3$) transformation of the $\psi^{A}_p$ coordinates, the total contribution to the tangent space coordinate does not change. 

The $2-$forms being dual to the 1-forms, show similar properties under any rotation in the internal $S^3$ space and the dot products i.e. $\psi^{*A}_{12}\psi^{B}_{12}+\psi^{*A}_{23}\psi^{B}_{23}+\psi^{*A}_{31}\psi^{B}_{31}$ remains constant under any SU($3$) transformation. The $3-$form is dual to the $0^{th}$ form and is a volume form. Therefore, it does not transform under rotation between the basis vectors $e^p$. The tangent space coordinate comes from the dot product as $X^a=\mathcal{G}^a_{AB}\psi^{A}_{123}\psi^{*B}_{123}$.

\subsection{\label{subsectionSU3}Rotation of {$\Omega_1$} and {$\Omega_2$} in complex coordinates}

If we rotate the coordinate system in the internal space, then the basis vectors $e^p$ change to, let us say, $e'^q$. These can be related as $e'^q = \mathcal{R}^q_p e^p$, where $\mathcal{R}$ is a complex rotation matrix and $\mathcal{R}^q_p$ is the element in the $p^{\text{th}}$ row and  $q^{\text{th}}$ column of the matrix. In $S^3$, it is a $3\times 3$ matrix whose elements can be complex numbers. So, the rotation along the complex axis in a single direction, which is prohibited in real rotation like SO($3$), is allowed in the complex rotation matrix ( check appendix~\ref{appendixC} for proper visualization). For SU($3$) rotation the matrix must satisfy $\mathcal{R}^T\mathcal{R} = 1$ and $\det(\mathcal{R})=1$. For small rotation, we can write the SU($3$) rotation matrix as $\exp(i\sum_{i=1}^8\omega_i \tau_i)$, where $\tau_i$ for $i=1, \cdots, 8$ are the generators of the SU($3$) group. The rotation along these generators is given by $\omega_i$s respectively.

Under SU($3$) transformation the differential 1-form transforms as ${\psi'}_p^A=\mathcal{R}^q_p{\psi}_q^A$, which in matrix format we can write as

\begin{equation}
\left[{\psi'}_p^A\right] = \left[\exp(i\sum_{i=1}^8\omega_i \tau_i) \right] \left[\psi_p^A\right]\,.\qquad
\end{equation}

\noindent Under covariant differentiation, $\psi^A_{p}$ have two connection parameters, one corresponding to the spin index $A$ and the other to the gauge index $p$. Thus the covariant differentiation can be written as 

\begin{equation}
\psi^A_{p;\mu} = \frac{\partial \psi^A_{p}}{\partial x^\mu} - \Gamma^q_{p\mu}\psi^A_{q} + \Gamma^A_{B\mu}\psi^B_{p} \,.
\end{equation}

\noindent The transformation rules for the $\Gamma^{A}_{B\mu}$ are given by Eq.~\ref{Gamma_matrix_spin_transform}. $\Gamma^{p}_{q\mu}$ is the connection parameter for the gauge variable. If we rotate the basis vectors of the internal coordinate system by a rotation matrix $\mathcal{R}^p_q$, we can write $\psi'^A_{p;\mu} = \mathcal{R}^q_p \psi^A_{q;\mu}$ which gives  

\begin{eqnarray}
\psi'^A_{p;\mu} &=& \frac{\partial \left(\mathcal{R}^q_p \psi^A_{q}\right)}{\partial x^\mu} - \Gamma'^s_{p\mu} \left(\mathcal{R}^q_s \psi^A_{q}\right) =  \mathcal{R}^q_p\frac{\partial \psi^A_{q}}{\partial x^\mu} 
- \left(\Gamma'^t_{r\mu}\mathcal{R}^s_t        
- \frac{\partial \mathcal{R}^s_r}{\partial x^\mu}\right)\psi^A_{s}
\mathcal{R}^q_p\mathcal{R}^v_w\delta^{rw}\delta_{qv} \nonumber\\
 &=& \mathcal{R}^q_p\left[\frac{\partial \psi^A_{q}}{\partial x^\mu} 
- \left(\Gamma'^t_{r\mu}\mathcal{R}^s_t        
- \frac{\partial \mathcal{R}^s_r}{\partial x^\mu}\right)\psi^A_{s}
\mathcal{R}^v_w\delta^{rw}\delta_{qv}\right]
= \mathcal{R}^q_p\left[\frac{\partial \psi^A_{q}}{\partial x^\mu}
- \Gamma^s_{q\mu}\psi^A_{s}\right] = \mathcal{R}^q_p \psi^A_{q;\mu}\,.
\end{eqnarray}

\noindent $\mathcal{R}$ being rotation matrix $\mathcal{R}\mathcal{R}^T = 1$, giving $\mathcal{R}^u_p\mathcal{R}^v_q\delta^{rq}\delta_{uv} = \delta^r_p$, where $\delta_{pq}=\delta^{pq}=\delta^p_q$ are the Kroniker deltas, i.e. $\delta^{pq} = \delta_{pq} = \delta^p_q = 1$ when $p=q$, and $0$ otherwise. This simplifies the transformation rules for the gauge connection parameter, and simple algebraic manipulation shows that $\Gamma^t_{r\mu}$ under internal basis rotation transforms as   

\begin{eqnarray}
 \Gamma'^s_{q\mu} = \left(\Gamma^t_{r\mu}\mathcal{R}^s_t        
+ \frac{\partial \mathcal{R}^s_r}{\partial x^\mu}\right)
\mathcal{R}^v_w\delta^{rw}\delta_{qv}\,.
\label{Internal_gamma_transformation}
\end{eqnarray}

Under the space-time, coordinate $x^\mu$ transformation, each $(s,q)$ component of $\Gamma^s_{q\mu}$ transforms as a normal vector as there is only one space-time index. For SU($3$) transformation, we can write the connection parameter  $\Gamma^s_{q\mu}$ as a sum of 8 vectors as 
$\Gamma^s_{q\mu}= i\sum_{i=1}^8 A^i_\mu[\tau_i]^s_q)$. Here $\mathbf{\tau}_i$s are the generators of the SU($3$) matrix and the $[\tau_i]^s_q$ is the ($s, q$) element of the matrix $\mathbf{\tau}_i$. $A^i_\mu = \frac{\partial \omega^i}{\partial x^\mu}$, is a vector field. 

In the internal space, the 2-forms are dual to the 1-forms and transform as SU($3$).
For the two form $\psi^A_{pq} = 0$ if $p=q$ and $\psi^A_{pq} = - \psi^A_{qp}$. For simplifying, if we define $\varphi^{Ar} = \psi^A_{pq}\varepsilon^{pqr}$ and $\varphi'^{Ar} = \psi'^A_{pq}\varepsilon^{pqr}$, then under the internal rotation $\varphi^{Ar}$ transforms as $\varphi'^{Ap}=\mathcal{R}^p_q\varphi^{Aq}$. Here, $\varepsilon^{pqr}$ are Levi-Civita symbol, $\varepsilon^{123} = \varepsilon^{231} = \varepsilon^{312} = 1$, $\varepsilon^{321} = \varepsilon^{132} = \varepsilon^{213} = -1$ otherwise $\varepsilon^{...} = 0$. In matrix format, we can write this transformation as 

\begin{equation}
\left[{\varphi'}^{Ar}\right] = \left[\exp(i\sum_{i=1}^8\omega_i \tau_i)\right] \left[\varphi^{Ar}\right]\,.
\end{equation}

\noindent Here also the covariant derivative of $\varphi^{Ap}$ is given by the same connection parameter as 

\begin{equation}
\varphi^{Ap}_{;\mu} = \frac{\partial \varphi^{Ap}}{\partial x^\mu} + \Gamma^p_{q\mu}\varphi^{Aq} + \Gamma^A_{B\mu}\varphi^{Bp}      \,.
\end{equation}

\noindent The 0-form does not change under any coordinate transformation. The 3-form is the volume form in this case and does not change under the rotation of the coordinates.

\subsection{Unitary transformation of the coordinates}

Apart from the rotational transformation SU($3$), we can also have the unitary U($1$) transformation in the complex space. As $\psi^A_{...}$ are the complex numbers, every coordinate axis has its own submanifold, which gives a U($1$) transformation. Under U($1$) gauge translation, the basis transforms as $e'^p=\exp(i\alpha)e^p$, where $\alpha$ is a small rotation angle. In matrix format, we can write this as 

\begin{equation}
\left[\begin{array}{c}
e'^{1} \\
e'^{2} \\
e'^{3} 
\end{array}\right]
= 
\left[\begin{array}{ccc}
\exp(i\alpha) & 0 & 0\\
0 & \exp(i\alpha) & 0\\
0 & 0 & \exp(i\alpha)
\end{array}\right]
\left[\begin{array}{c}
e^{1} \\
e^{2} \\
e^{3} \\
\end{array}\right]      \,.
\label{rotation_and_translation_matrix}
\end{equation}

\noindent Under this transformation, different forms are shown in Eq.~\ref{diffenrential_forms_series_1} transform as  

\begin{alignat}{2}
&\Omega'_0 : \psi'^A  && \qquad\, \quad \\
&\Omega'_1 : \psi'^A_p\exp(i\alpha) e^p  && \qquad\cdots \quad\forall p\in (1,2,3)\\
&\Omega'_2 : \psi'^A_{pq}\exp(i2\alpha) e^p\wedge e^q   && \qquad\cdots \quad\forall p, q\in (1,2,3) \;\;\;\;\;\;\&\; p, q\, \text{in circular order}\label{2formtransformation}\\
&\Omega'_3 : \psi'^A_{pqr}\exp(i3\alpha) e^p\wedge e^q\wedge e^r  && \qquad\cdots \quad\forall  p, q, r\in (1,2,3)\;\;\;\&\; p<q<r
\label{DiffForm12}
\end{alignat}

\noindent Therefore, the 0-form remains unchanged, and $\psi'^A = \psi^A$. As in the internal $S^3$ space $\psi^A_p$ is a vector, it should not change under any coordinate transformation, and only its components along different basis vector changes, giving $\psi'^A_p e'^p = \psi^A_p e^p$. Hence, the 1-forms transform as 

\begin{equation}
\psi'^A_p = \exp(-i\alpha)\psi^A_p  \,.
\end{equation}

\noindent Here one should note that $\exp(i\alpha)$ is $\frac{1}{3}$ of the tress of the $3\times 3$ matrix given in Eq.~\ref{rotation_and_translation_matrix}. The connection parameter for the covariant derivative for this transformation is also similar to the connection parameter for the SU($3$) rotation, except it only affects the $\Gamma^p_{p\mu}$ parameters.   
If we do not take any changes in the spin coordinate, then the covariant derivative of $\psi'^A_p$ and $\psi^A_p$ can be related as 

\begin{eqnarray}
\psi'^A_{p;\mu} &=& \exp(-i\alpha)\frac{\partial\psi^A_p}{\partial x^\mu} - i\frac{\partial \alpha}{\partial x^\mu}\exp(-i\alpha)\psi^A_{p} -\Gamma'^p_{p\mu}\exp(-i\alpha)\psi^A_{p} \nonumber  \\
&=& \exp(-i\alpha)\left[\frac{\partial\psi^A_p}{\partial x^\mu} - \left(\Gamma'^p_{p\mu}+i\frac{\partial \alpha}{\partial x^\mu} \right)\psi^A_{p}\right]    =
\exp(-i\alpha)\left[\frac{\partial\psi^A_p}{\partial x^\mu} - \Gamma^p_{p\mu}\psi^A_{p}\right]\,.  \end{eqnarray}

\noindent Hence, under U($1$) transformation the connection parameter, i.e. $\Gamma^p_{p\mu}$, transforms as $\Gamma'^p_{p\mu}=\Gamma^p_{p\mu}-i\frac{\partial \alpha}{\partial x^\mu}$. Note that there is no summation over $p$. If we assume no SU($3$) gauge transformation and only U($1$) gauge transformation, then for such transformation, we can consider the connection parameters $\Gamma^1_{1\mu} = \Gamma^2_{2\mu} = \Gamma^3_{3\mu} = iA_\mu$, where $A_\mu = \frac{\partial \alpha}{\partial x^\mu}$ is an arbitrary vector field and all the other $\Gamma^p_{q\mu}$ are $0$. 

Similarly, under U($1$) transformation, the two forms show satisfy $\psi'^A_{pq}e'^p\wedge e'^q = \psi^A_{pq}e^p\wedge e^q$, giving $\psi'^A_{pq} = \exp(-2i\alpha)\psi^A_{pq}$ for all $p,q\in(1,2,3)$ and $p$, $q$ in circular order as shown in Eq.~\ref{2formtransformation}. The 3-form transforms as $\psi'^A_{123} = \exp(-3i\alpha)\psi^A_{123}$. The covariant derivative for the 2-form and 3-form can be written as 

\begin{equation}
\psi^A_{pq;\mu} = \frac{\partial \psi^A_{pq}}{\partial x^\mu} -\Gamma^p_{p\mu}\psi^A_{pq}-\Gamma^q_{q\mu}\psi^A_{pq}=\frac{\partial \psi^A_{pq}}{\partial x^\mu} -2iA_\mu \psi^A_{pq}\,,
\end{equation}
\noindent and
\begin{equation}
\psi^A_{pqr;\mu} = \frac{\partial \psi^A_{pqr}}{\partial x^\mu} -\Gamma^p_{p\mu}\psi^A_{pqr}-\Gamma^q_{q\mu}\psi^A_{pqr}-\Gamma^r_{r\mu}\psi^A_{pqr}=\frac{\partial \psi^A_{pqr}}{\partial x^\mu} -3iA_\mu \psi^A_{pqr}\,.
\end{equation}

\noindent Here, one can see that the coupling strength of the coupling parameter for the 2-form is two times, and the 3-form is three times that of the 1-form.

\subsection{Renaming the coordinate systems}
In the previous section, we define total $8$ sets of parametric space coordinate systems. Each coordinate system can represent a unique point in the tangent space. In these coordinate systems, there are two triplets, the 1-forms, and the 2-forms. These are sensitive to the rotation in the internal coordinate system. Therefore, if a particle, represented by these forms, moves freely from one point to another in space-time, then the phase of the parametric coordinate system saves the information about all the rotations in each direction in the internal $S^3$ space through its path.  

We can associate these eight coordinates to represent different Fermions. The 0-form neither couples with any U($1$) field nor any SU($3$) field. Therefore, we can use it to represent a neutrino or an anti-neutrino. The 3-form does not couple with any SU($3$) field; however couples with a U($1$) field. Hence, it can be used to represent an electron or a position. The 1-forms couple with an U($1$) field with $\frac{1}{3}$ strength of that of the 3-forms. Therefore, it can be used to represent a down quark or an anti-down quark. Each 1-form has three components which can be rotated as SU($3$) group. They represent 3 color charges of the d-quark triplet ($\Psi_{d_r}^A$, $\Psi_{d_g}^A$, $\Psi_{d_b}^A$). The SU($3$) connection parameters represent $8$ gluons. The 2-forms can be used to represent the up or the anti-up quarks as they couple with a U($1$) field with a $\frac{2}{3}$ strength of that of the 3-forms. Also, they couple with the SU($3$) field. Therefore, we can rename the parametric space coordinate systems as

\noindent\begin{minipage}{0.30\columnwidth }
\begin{eqnarray}
\Psi_{\nu}^A &=&  \psi^A(x_i)\,, \nonumber \\
\Psi_{e}^A &=& \psi_{123}^A(x_i)\,, \nonumber 
\end{eqnarray} 
\end{minipage}
\begin{minipage}{0.33\columnwidth }
\begin{eqnarray}
\Psi_{d_r}^A &=& \psi_1^A(x_i)\,, \nonumber \\
\Psi_{d_g}^A &=& \psi_2^A(x_i)\,, \nonumber \\
\Psi_{d_b}^A &=& \psi_3^A(x_i)\,, \nonumber
\end{eqnarray} 
\end{minipage}
\begin{minipage}{0.33\columnwidth }
\begin{eqnarray}
\Psi_{u_r}^A &=& \psi_{23}^A(x_i)\,, \nonumber \\
\Psi_{u_g}^A &=& \psi_{31}^A(x_i)\,, \nonumber \\
\Psi_{u_b}^A &=& \psi_{12}^A(x_i)\,. \label{Gauge_Equation_set_1}
\end{eqnarray} 
\end{minipage}
\; \newline

The spin indices represent the spin or the right or left-handedness of the particles (for massless particles, helicity and chirality are the same). This article uses the four-dimensional space-time to generate parametric space coordinates to represent the massless particles. As the spinors in the article are massless, it is not possible to distinguish between the particle and the antiparticles.
However, according to our previous articles, the space-time should be $5$ dimensional. In a higher dimension, we can introduce a similar parametric space coordinate system with mass terms and get separate coordinates for particles and antiparticles. However, this discussion is beyond the scope of this article and will be addressed in a future article. We also need to develop theories of weak interaction, i.e., the SU($2$) transformation between the electrons and the neutrinos. We want the doublets ($\Psi^A_\nu, \Psi^A_e$), ($\Psi^A_u, \Psi^A_d$) to transform as a SU($2$) system. This issue will also be addressed in future articles.

\section{Discussion and Conclusion}

In this paper, we explore the motion of a massless particle moving freely on a null hyper-surface of a 4-dimensional space-time manifold. We define a tangent space on every point of the manifold. For a massless particle, the tangent space coordinates represent the space of the 4-velocity of the particle. We define a complex parametric space coordinate system on the tangent space. The parametric space coordinate system behaves like spinors. We show that the parametric space coordinates have an additional degree of freedom that can be used to store the path integral of a vector field through the path of the particle. We explore different properties of the parametric space coordinates. We define a 3-sphere ($S^3$) space on the tangent space and define eight differential forms from the parametric space coordinate system. These coordinate systems can couple with different SU($3$) and U($1$) fields. Therefore, they can be used to represent different elementary Fermions. In this paper, the exercise has been carried out on a 4-dimensional manifold that gives only massless spinors. However, it is possible to accommodate massive spinors in higher dimensions, which we plan to address in our subsequent research.

\appendix

\section{\label{Lp_space} Consequence in {$L^p$} space}
In an $L^2$ space, the length between any two points is given by the square root of the sum of the square of the distances between those two points in different directions, i.e., $d = [(X^0)^2 - \sum_{i=1}^{n-1} (X^i)^2]^\frac{1}{2}$ in an $n$ dimensional Minkowski spacetime. ( In a Minkowski spacetime, the $X^0$, i.e., the time component, comes up with a positive signature, and the rest of spatial dimensions appear with a negative signature.) In this paper, we take a null manifold in a 4-dimensional spacetime and define a parametric coordinate system on its tangent space. We show that these parametric coordinates can be assigned a tensorial weight of $\frac{1}{2}$, which we call spin weight. We also lower the spin indexes of our parametric coordinate system to get a system of spin $-\frac{1}{2}$ variables. 

Even though our space is an $L^2$ space, mathematically, we are free to consider an $L^p$ space where the distances in different dimensions add up to the power of $p$. In such a case, for a $n$ dimensional space, the distance element is given by $d = [(X^0)^p - \sum_{i=0}^{n-1} (X^i)^p]^\frac{1}{p}$. Here $X^i$ represent the distance between two points along $i^{th}$ direction and it is assumed to be a real positive quantity. In case of a negative $X^i$ we should use its modulus. 

For null manifold on a 4-dimensional $L^p$ space, the tangent space is given by 

\begin{equation}
    (X^0)^p - (X^1)^p - (X^2)^p - (X^3)^p = 0\,.
\end{equation}

\noindent Here, we consider the spatial coordinates to have a negative signature as a general convention. However, it does not affect any calculation. If we consider $\omega_p = -\exp(2i\pi/p)$ as the $p^{\text{th}}$ root of $-1$, then we can factorize the above equation as
\begin{equation}
(X^0 - X^1)(X^0 + \omega_p X^1)\ldots(X^0 + \omega^{p-1}_p X^1) = (X^2 + X^3)(X^2 - \omega_p X^3)\ldots(X^2 - \omega^{p-1}_p X^3)\,,
\end{equation}

\noindent and follow the logic of Sec.~\ref{sectionTangentSpace}, to parameterize the tangent space. For simplicity, let us consider an $L^3$ space and break it as 
\begin{eqnarray}
(X^0 - X^1)(X^0 + \omega X^1)(X^0 + \omega^{2} X^1) = (X^2 + X^3)(X^2 - \omega X^3)(X^2 - \omega^{2} X^3)\,,
\label{A3cubic}
\end{eqnarray}

\noindent where $\omega$ is the imaginary cube-root of $-1$. We can write it as a system of three linear equations 
\begin{eqnarray}
 X^1+X^2 &=& \lambda_1 (X^0-X^3)\,, \\
 X^1-\omega X^2 &=& \lambda_2 (X^0 + \omega X^3)\,,\\
 X^1-\omega^2 X^2 &=& \frac{1}{\lambda_1 \lambda_2} (X^0 + \omega^2 X^3)\,.
\end{eqnarray}

\noindent Here $\lambda_1$ and $\lambda_2$ are two arbitrary numbers. $\lambda_1$ has to be real as the coordinates are real numbers. However, $\lambda_2$ can be a real or complex number. We can permute the combinations in the left and right-hand sides of the equations to get different systems of linear equations. The parameterization of the tangent space can be done as 
\begin{eqnarray}
 X^1+X^2 &= \varPsi^1 \varPsi^{*1}\,, \qquad\qquad X^0-X^3 &= \varPsi^2 \varPsi^{*2}\,, \nonumber \\
 X^1-\omega X^2 &= \varPsi^3 \varPsi^{*2}\,, \qquad\qquad X^0 + \omega X^3 &= \varPsi^3 \varPsi^{*1}\,, \nonumber \\
 X^1-\omega^2 X^2 &= \varPsi^{*3} \varPsi^2\,, \qquad\qquad X^0 + \omega^2 X^3 &= \varPsi^{*3} \varPsi^1\,.
 \label{EqA7}
\end{eqnarray}

\noindent As we can see, we need $3$ complex numbers, $\varPsi^A$  for $A\in(1,2,3)$ to parameterize this space. In fact, similar calculations can show that we need total $p$ complex numbers to parametrize the tangent space of an $L^p$ space. As there are only $3$ degrees of freedom in the tangent space, the parametric space has $2p-3$ additional degrees of freedom, which can be used as gauge freedom. As before, we refer to $A\in(1,\dots,p)$ as the spin index.

Under linear transformation, $\varPsi^A$, transforms as $\bar{\varPsi}^A = T^A_B \varPsi^B$, which we call the spin space transformation. This linear transformation also satisfies the cubic equation Eq.~\ref{A3cubic}, though the equation gets multiplied with $T^2$ where $T=|T^A_B|$.
For giving a tensorial structure to these parametric space coordinates, let us define $\varPsi_A$ as the covariant counterpart of $\varPsi^A$. Suppose we use some quantity $\epsilon_{AB}$ to lower the index of $\varPsi^A$. 
This gives 

\begin{eqnarray}
&\epsilon_{AB}\varPsi^A\varPsi^B = \bar{\epsilon}_{CD}\bar{\varPsi}^C\bar{\varPsi}^D = \bar{\epsilon}_{CD}T^C_BT^D_B\varPsi^A\varPsi^B \,,\\   
&\Rightarrow  \epsilon_{AB} = \bar{\epsilon}_{CD}T^C_BT^D_B   \,.
 \label{Spinspacerot3D}
\end{eqnarray}

\noindent This may be a good choice for raising or lowering the indices, but if we demand the $\epsilon_{AB}$ to remain unchanged under coordinate transformation, then mathematically, it is impossible to make $\epsilon_{AB}$ and $\bar{\epsilon}_{CD}$ equal. As the parametric space is a 3-dimensional space, $\epsilon_{AB}$ is a two-form, and it has some direction in the parametric space. When we make a linear transformation of the parametric coordinates, it changes the direction of $\epsilon_{AB}$ in the parametric space. 

If we want a quantity that does not change under such linear transform, we need to have a 3-form or a volume form, e.g., $\epsilon_{ABC}$. Consequently, in $L^p$ space where we have $p$ parametric space coordinates, we need a $p$-form that can remain unaltered under a coordinate transformation. 

Therefore, in 3-dimensional space if we want a  constant quantity to raise or lower the indices we should take 

\begin{eqnarray}
&\epsilon_{ABC}\varPsi^A\varPsi^B\varPsi^C = \bar{\epsilon}_{DEF}\bar{\varPsi}^D\bar{\varPsi}^E\bar{\varPsi}^F = \bar{\epsilon}_{DEF}T^D_AT^E_BT^F_C\varPsi^A\varPsi^B\varPsi^C \,,  \\
&\Rightarrow  \epsilon_{ABC} = \bar{\epsilon}_{DEF}T^D_AT^E_BT^F_C \,.  
\label{EqA11}
\end{eqnarray}

Provided we choose $\epsilon_{ABC}$ to be the Lavi-Civita symbol in three dimensions, then based on the properties of the Lavi-Civita symbol, this transformation multiplies the left-hand side of Eq.~\ref{EqA11} by the determinant of $T^A_B$. To keep the $\epsilon_{ABC}$ same in both frames, we need to multiply the right side by $T^{-1}$, where $T=|T^A_B|$. Thus the transformation looks like

\begin{equation}
\epsilon_{ABC} = T^{-1}\epsilon_{DEF}T^D_AT^E_BT^F_C    \,.
\end{equation}

\noindent Lowering or raising indices using $\epsilon_{ABC}$ and $\epsilon^{ABC}$ takes the form 

\begin{eqnarray}
\varPsi_{BC} = \epsilon_{ABC}\varPsi^A\,, \qquad \varPsi_{BC} = \epsilon_{ABC}\varPsi^A\,, \qquad  \varPsi^{AB}=\epsilon^{ABC}\varPsi_C\,, \qquad \varPsi^{A}=\epsilon^{ABC}\varPsi_{BC}\,.
\end{eqnarray}

\noindent It is essential to note that the Levi-Civita symbol is not a symmetric quantity. The indices at the right are used to lower an index, and those at the left are used to raise an index. Also, in this setup when we raise or lower an index we get two indices at the top or bottom respectively. 

Here $\epsilon^{ABC}$ and $\epsilon_{ABC}$ behave as a  contravariant and covariant tensor density with weight $+1$ and $-1$, respectively. Raising indices with $\epsilon^{ABC}$ and lowering indices with $\epsilon_{ABC}$ increases and decreases the weight of a quantity by $+1$ and $-1$. 
Therefore, if we assign a spin weight of $\frac{1}{3}$ to each of the $\varPsi^A$ coordinates,
then $\varPsi_{AB}$ should have a spin weight $-\frac{2}{3}$. $\varPsi_A$ and $\varPsi^{AB}$ get spin weight $-\frac{1}{3}$ and $\frac{2}{3}$ respectively. In such case, under spin space transformation, the $\varPsi$-coordinates transform as

\begin{equation}
    \bar{\varPsi}^A = T^{-\frac{1}{3}}T^A_B\varPsi^B. 
\end{equation}

\noindent This transformation has a determinant 1. Therefore, if we introduce it back to the cubic polynomial, given in Eq.~\ref{A3cubic}, it does not multiply the equation by $T^2$. 

This concept can be generalized for any $L^{p}$ spaces, and we introduce a spin $\frac{1}{p}$ parametric space coordinate system in $L^p$ spaces. We need $p$ components of $\varPsi^A$ to span the space and need a $p$ dimensional Lavi-Civita symbol for raising and lowering the spin indices. However, these parametric space coordinate systems are not unique, especially when $p$ is not a prime number. For instance, if $p=6$, then we can introduce a spin $\frac{1}{2}$ or spin $\frac{1}{3}$ coordinate system instead of a spin $\frac{1}{6}$ coordinate system and these coordinate systems can span the space equally well. However, by doing so, we lose some gauge freedom in the system.

\section{\label{appendixB}Understanding the rotation in the internal coordinate system}

To understand the rotations in the internal coordinate system, we consider a 3-dimensional spatial manifold. The coordinate system on this manifold is given by $x^\mu$ for $\mu\in(1,3)$, and the line element is  $ds^2=g_{\mu\nu}dx^\mu dx^\nu$. Let us take a car moving through a geodesic path on this manifold, and its velocity is given by $v^\mu = \frac{dx^\mu}{ds}$. Suppose there is a coordinate system, $X^a$, that is attached to the car. These $X^a$s represent the tangent space coordinate system, and we can relate these to the velocity of the car as $X^a = \Lambda^a_\mu \frac{dx^\mu}{ds}$.

Let us consider that a setup is installed on the car that can measure the velocity of the wind coming from three different directions and rotate an arrow. The arrow should point along the direction of the wind. This setup introduces three extra degrees of freedom or three new coordinates. We can consider the basis vectors for this coordinate system as $e^p$, and the rotation of the arrow is measured with respect to these basis vectors. The arrow can only rotate with respect to these basis vectors and has no translational or any other kind of motion with respect to $e^p$. This rotational space represents the internal space of our analysis. Note that, for visualization purposes, we are working on real space, and hence we can always align the basis $e^p$ to coincide with the tangent space coordinate axis, $X^a$. However, in reality, that is impossible as these basis vectors and the tangent space denote two distinct hyperspaces. 

If we assume that $\vec{V}$ represents the direction of the arrow on this setup, then the vector $\vec{V}$ should be measured with respect to the local basis vector $e^p$. The arrow is always pointed towards the direction of the wind. So if we look from the wind's reference, the arrow is fixed, but the local basis vectors are changing. We can write its covariant derivative with respect to the spacetime coordinate as

\begin{equation}
    V^{p}_{;\nu} = \frac{\partial V^{p}}{\partial x^\nu} + \Gamma^p_{q \nu} V^{q} \,.
    \label{ArrowcovariantDerivative}
\end{equation}

\noindent Here $\Gamma^p_{q \nu}$ are the connection parameters for representing the rotation of the arrow with respect to the basis vectors $e^p$. $\Gamma^p_{q \nu}$ has only one spatial index. Therefore, essentially it is the sum of some vector fields.


Now, if we make any transformation in the basis vectors from $e^p \rightarrow e'^p$, then the $V^{p}$ must transform as

\begin{equation}
    V'^{q} = V^{p}\frac{\partial e'^q}{\partial e^p}
    \qquad\Rightarrow\qquad
    \frac{\partial V'^{q}}{\partial x^{\nu}} = \frac{\partial V^{p}}{\partial x^{\nu}}\frac{\partial e'^q}{\partial e^p} + V^{ p}\frac{\partial^2 e'^q}{\partial x^{\nu} \partial e^p}\,.
    \label{CoordinateTrndformTensor}
\end{equation}



As we are working in a 3D real space, any general rotation of the basis is given by SO($3$) transformation, which has three generators 
\begin{equation}
\tau_1 = \left[\begin{array}{ccc}
0 & 0 & 0 \\
0 & 0 & -1 \\
0 & 1 & 0
\end{array}\right], \quad \tau_2 = \left[\begin{array}{ccc}
0 & 0 & 1 \\
0 & 0 & 0 \\
-1 & 0 & 0
\end{array}\right], \quad \tau_3 = \left[\begin{array}{ccc}
0 & -1 & 0 \\
1 & 0 & 0 \\
0 & 0 & 0
\end{array}\right]\,. \nonumber
\end{equation}

\noindent Therefore, in general, for small rotation $\theta_1$, $\theta_2$ and $\theta_3$ of the coordinate system along the basis vectors $e^1$, $e^2$ and $e^3$, the new basis vectors are given by 
\begin{equation}
\begin{bmatrix}
e'^1 \\
e'^2 \\
e'^3 
\end{bmatrix}
= \exp\left(
\theta_1 \left[\begin{array}{ccc}
0 & 0 & 0 \\
0 & 0 & -1 \\
0 & 1 & 0
\end{array}\right] + \theta_2 \left[\begin{array}{ccc}
0 & 0 & 1 \\
0 & 0 & 0 \\
-1 & 0 & 0
\end{array}\right] + \theta_3 \left[\begin{array}{ccc}
0 & -1 & 0 \\
1 & 0 & 0 \\
0 & 0 & 0
\end{array}\right]
\right)\begin{bmatrix}
e^1 \\
e^2 \\
e^3 
\end{bmatrix}\,.
\end{equation}

\noindent Therefore, the second term of the Eq.~\ref{CoordinateTrndformTensor}, can be written in the matrix format as  
\begin{equation}
\left[\frac{\partial^2 e'^p}{\partial x^\nu \partial e^q}\right]
= \left(\sum_{i=1}^3
\frac{\partial \theta_i}{\partial x^\nu} \tau_i \right) \exp\left(\sum_{i=1}^3
\theta_i \tau_i \right) \,.
\label{doublederivative_XXx}
\end{equation}

\noindent In Eq.~\ref{ArrowcovariantDerivative}, if the covariant derivatives come only from the rotation of the arrow, then $V^{p}_{;\nu} = \Gamma^p_{q\nu} V^{q}$. If the basis vectors of the rotational axis keep changing over the space, then with respect to those basis vectors, $V^{p}$ also change as 
$V'^{p} = \left[\exp(-\sum_{i=1}^{3} \theta_i \tau_i)\right]^p_q V^{q}$, where $[\dots]^p_q$ is the $p,q$th term of the matrix.  Therefore, the conformal derivative of $\vec{V'}$ is given by 
\begin{equation}
V'^{p}_{;\nu} = \left(\Gamma^p_{q\nu} - \left[\sum_{i=1}^3\frac{\partial \theta_i}{\partial x^\nu} \tau_i \right]^p_q\right) \left[\exp(-\sum_{i=1}^{3} \theta_i \tau_i)\right]^q_r V'^{ r}_{;\nu}\,.
\end{equation}

\noindent As discussed before, $\Gamma^p_{q\nu}$ only has a single spatial index. Therefore it behaves as a linear combination of vectors. If we consider $\Gamma^p_{q\nu} = \sum_{i=1}^3 W_{i\nu} \tau_i$ then it simplifies the equations, where $W_{1\nu}$, $W_{2\nu}$ and $W_{3\nu}$ are three vector fields. In this case $W_{i\nu}$ transforms as $W'_{i\nu} = W_{i\nu}-\frac{\partial \theta_i}{\partial x^\nu}$, which is equivalent to the expression shown in Eq.~\ref{Internal_gamma_transformation}. 

Suppose some wind blows through the path; it changes the direction of the arrow. $W_{p\nu}$ can be thought as some torque basis vectors. So, as the car moves from point $\mathcal{P}$ to point $\mathcal{Q}$ through different paths, it changes the direction of the arrow based on the path. The final direction of the arrow will be different for two different paths.

For checking the dependence on the path of the car, multiply Eq.~\ref{ArrowcovariantDerivative} with $\frac{dx^\mu}{ds}$ to get 
\begin{equation}
\frac{dV^{p}}{ds} = V^{ p}_{;\nu}\frac{dx^\nu}{ds} = \frac{\partial V^{ p}}{\partial x^\nu}\frac{dx^\nu}{ds} + \left[\sum_{i=1}^3 W_{i\nu} \tau_i\right]^p_q V^{q}\frac{dx^\nu}{ds} \,.
\end{equation}

\noindent Here, $s$ is a line element through the path, which is an affine parameter. When the car moves from point $\mathcal{P}$ to $\mathcal{Q}$, the total change in the direction of $V^{q}$ from the wind is $\int_\mathcal{P}^\mathcal{Q}  \left[\sum_{i=1}^3 W_{i\nu} \tau_i\right]^p_q V^{q}\frac{dx^\nu}{ds} ds$. Therefore, the final $V^{p}$ stores the information of the external vector field through the path. 

In this example, the rotation of the arrow vector with respect to the basis vectors is given by a SO($3$) rotation because the arrow vector is a real vector. However, for complex vectors in the internal space as discussed in Sec.~\ref{sec6}, this rotation is given by a SU($3$) rotation group, which has total $8$ independent $\theta$s and the connection parameter is linked to $8$ independent vector fields. We can call these vector fields gluon fields. When some particle, denoted by parametric space coordinate system, $\psi^A_p$, moves through these vector fields, it can rotate in the internal coordinate space. We can also think of it as if the local internal coordinate systems are changing. So as the particle is moving its component along different direction are changing. The component of $\psi^A_p$ along different basis vectors in the internal coordinate system gives the component of $\psi^A_p$ along different color charges. The final direction of the $\psi^A_p$ in the internal coordinate system stores the information of the path integral with the eight external fields.

\section{\label{appendixC}Visualizing rotation on a complex manifold}

\subsection{A setup for visualizing complex rotations}
In this section, we discuss the SU($3$) transformations for the complex coordinates~\cite {Coddens2018RenderingSI}. Even though the generators of these rotation groups can be determined mathematically, it is complicated to visualize the complex rotations. Therefore, here we try to provide an intuitive understanding of the complex rotations, which may help the readers to comprehend the rotations discussed in the paper, especially in Sec.~\ref{subsectionSU3}. 

For visualizing the complex rotations, we take an example of an electromagnetic field. It has an electric and a magnetic component, given by $\vec{E} = (E_x, E_y, E_z)$ and $\vec{B} = (B_x, B_y, Bz)$ respectively. The total energy density of the field can be written as $\mathcal{E}^2=\epsilon_0 \bigg(E_x^2+E_y^2+E_z^2+B_x^2+B_y^2+B_z^2\bigg)$, where $\epsilon_0$ is the permittivity of vacuum, and we consider $c=1$. We define the field-strength along different direction as  $F_x = \sqrt{E_x^2+B_x^2}$, $F_y = \sqrt{E_y^2+B_y^2}$, $F_z = \sqrt{E_z^2+B_z^2}$. Here, $\mathcal{E}^2$ and $\vec{F}$ are real quantities. Therefore, under coordinate transform, we want $\mathcal{E}^2$ to remain unaltered and $\vec{F}$ to transform as a real vector, i.e., $\vec{F}$ must follow the rotation group SO($3$). Let us assume 
\begin{equation}
\mathcal{F}_x = E_x+iB_x\,, \qquad\qquad \mathcal{F}_y = E_y+iB_y\,, \qquad\qquad \mathcal{F}_z = E_z+iB_z\,,
\end{equation}

\noindent i.e. $F^2_{i} = \mathcal{F}_{i} \bar{\mathcal{F}}_{i},\; \forall i\in (x,y,z)$. We can treat the directions of the electric field and magnetic fields' directions as two independent directions. As in this setup, the $\mathcal{F}$ is a complex number, and $E$ and $B$ are the real and the imaginary components of a complex number. We have a total of $6$ independent coordinates, which form the basis of the electromagnetic field.

First, let us keep the $E_z$ and $B_z$ components constant and consider the rotation between the X-Y plane only. A $2D$ complex space has a total of $4$ real coordinates with some additional structure. If there is a $4D$ real space, then we can rotate the coordinate system using SO($4$), which is easy to visualize. However, all the SO($4$) rotations can not be independent SU($2$) rotations. For example, in the X-Y plane, the $4$ basis vectors can be rotated in a total of 6 ways per SO($4$) rotation. However, the constraint is that under any type of rotation in the X-Y plane, the $\vec{F}$  must rotate as SO($2$) in the X-Y plane. If we make any rotation between $E_x$ and $B_y$, which is permitted under SO($4$), then $\vec{F}$ does not transform in the proper way unless we rotate some other axes in such a way that they restore the rotation of $\vec{F}$. Hence, the transformations under SU($2$) are constrained.

\subsection{Understanding the SU(3) rotation}
A rotation matrix $\mathcal{R}\in \textrm{U}(3)$ if it satisfies the condition $\mathcal{R}\mathcal{R}^T = \mathbf{1}$, where $\mathbf{1}$ is the unity matrix. To belong to SU($3$), it must also satisfy the condition $\det \mathcal{R} = 1$. There can be three forms of rotation in SU($3$) or any SU($n$). 

For understanding the first kind of rotation, let us rotate the complex field in the $x$ direction in the complex plane. Here we are not rotating the vector in the $X$,$Y$, and $Z$ axes. Instead, we rotate it in the complex $X$ plane, keeping the $Y$ and $Y$ planes fixed. We consider the initial fields as

\begin{eqnarray}
E_x = F_x\cos(\chi^z_0) \,,\qquad\qquad B_x = F_x\sin(\chi^z_0)\,.
\end{eqnarray}

\noindent After rotating the fields by $\delta\chi^z$ we have  $(E'_x+iB'_x) = (E_x+iB_x)\exp(i\delta\chi^z)$. Thus the rotation matrix for such transformation can be written as 

\begin{eqnarray}
\left[\begin{array}{c}
E'_{x}+\imath B'_{x} \\
E'_{y}+\imath B'_{y} \\
E'_{z}+\imath B'_{z} \\
\end{array}\right]
= 
\left[\begin{array}{ccc}
\exp(i\delta\chi^z) & 0 & 0\\
0 & 1 & 0\\
0 & 0 & 1
\end{array}\right]
\left[\begin{array}{c}
E_{x}+\imath B_{x} \\
E_{y}+\imath B_{y} \\
E_{z}+\imath B_{z} \\
\end{array}\right] \,.
\end{eqnarray}

\noindent Such rotation neither change $F_x$ nor the $\mathcal{E}$; hence, it is a valid rotation in the complex plane. Unfortunately, the determinant of the matrix is not unity, and hence it is a U($3$) transformation but not SU($3$). In fact, it is a combination of U($1$) and SU($3$). However, if we combine this with another rotation in the $Y$ direction, i.e.

\begin{equation}
\left[\begin{array}{ccc}
\exp(i\delta\chi^z) & 0 & 0\\
0 & \exp(-i\delta\chi^z) & 0\\
0 & 0 & 1
\end{array}\right]\,,
\end{equation}

\noindent then such transformation does comply with the unit determinant condition. We can use this technique with the other two axis pairs. The rotation matrices for those cases are
\begin{equation}
\left[\begin{array}{ccc}
1 & 0 & 0\\
0 & \exp(i\delta\chi^x) & 0\\
0 & 0 & \exp(-i\delta\chi^x)
\end{array}\right] \;,
\left[\begin{array}{ccc}
\exp(-i\delta\chi^y) & 0 & 0\\
0 & 1 & 0\\
0 & 0 & \exp(i\delta\chi^y)
\end{array}\right]\,.
\end{equation}

\noindent However, if we apply all three rotations together, then the complete rotation matrix will be 
\begin{equation}
\left[\begin{array}{ccc}
\exp(i\delta\chi^z-i\delta\chi^y) & 0 & 0\\
0 & \exp(i\delta\chi^x-i\delta\chi^z) & 0\\
0 & 0 & \exp(i\delta\chi^y-i\delta\chi^x)
\end{array}\right]
\label{center_rot}
\end{equation}

\noindent The interesting thing is to note that the sum of all the exponents is zero. It means that if we apply rotation $\delta\chi^z$ and $\delta\chi^y$ such that $\delta\chi^z -\delta\chi^y = 0$, then that gives the third rotation. In other words, there are only two degrees of freedom. The third rotation can be performed by a combination of the first two rotations. 
We can redefine the variables as 

\begin{eqnarray}
\delta\chi^y-\delta\chi^x &=& \delta \sigma_1  \\
\delta\chi^z-\delta\chi^x &=& \delta\chi^z- \delta\chi^y + \delta\chi^y - \delta\chi^x = \delta \sigma_2 + \delta\sigma_1 \\
\delta\chi^y-\delta\chi^z &=& -\delta \sigma_2
\end{eqnarray}

\noindent For small rotations Eq.~\ref{center_rot} gives us the two generators of SU($3$) as 

\begin{eqnarray}
\left[\begin{array}{ccc}
1 & 0 & 0\\
0 & 1 & 0\\
0 & 0 & 1
\end{array}\right]+
i\delta\sigma_2\left[\begin{array}{ccc}
1 & 0 & 0\\
0 & -1 & 0\\
0 & 0 & 0
\end{array}\right]+
i\delta\sigma_1\left[\begin{array}{ccc}
1 & 0 & 0\\
0 & -2 & 0\\
0 & 0 & 1
\end{array}\right]
\label{EMboost}
\end{eqnarray}

\noindent There are other kinds of rotations, i.e., the rotation between any two real coordinates. We want $\vec{F}$ to rotate as SO($3$). If $F_x$ and $F_y$ rotate by an angle $\theta_z$ along $Z$ axis then the transformation can be written as

\begin{eqnarray}
\left[\begin{array}{c}
F'_{x} \\
F'_{y} \\
F'_{z} \\
\end{array}\right]
= 
\left[\begin{array}{ccc}
\cos(\theta_z) & \sin(\theta_z) & 0\\
-\sin(\theta_z) & \cos(\theta_z) & 0\\
0 & 0 & 1
\end{array}\right]
\left[\begin{array}{c}
F_{x} \\
F_{y} \\
F_{z} \\
\end{array}\right]\,.
\label{Ftransformation}
\end{eqnarray}

\noindent Here $F_i$, $\forall i\in {(x,y,z)}$  are the absolute values of the complex electromagnetic field. This transformation can be done in $2$ ways. The simplest possibility is that both the $E$ and $B$ fields rotate in the same way, which gives us, 

\begin{eqnarray}
\left[\begin{array}{c}
E'_{x}+\imath B'_{x} \\
E'_{y}+\imath B'_{y} \\
E'_{z}+\imath B'_{z} \\
\end{array}\right]
= 
\left[\begin{array}{ccc}
\cos(\theta_z) & \sin(\theta_z) & 0\\
-\sin(\theta_z) & \cos(\theta_z) & 0\\
0 & 0 & 1
\end{array}\right]
\left[\begin{array}{c}
E_{x}+\imath B_{x} \\
E_{y}+\imath B_{y} \\
E_{z}+\imath B_{z} \\
\end{array}\right]\,.
\end{eqnarray}

\noindent Of course, it satisfies all the conditions required. For small values of $\theta_z$, we get 

\begin{eqnarray}
\delta \left[\begin{array}{c}
E_{x}+\imath B_{x} \\
E_{y}+\imath B_{y} \\
E_{z}+\imath B_{z} \\
\end{array}\right]
= 
\imath \theta_z\left[\begin{array}{ccc}
0 & -i & 0\\
i & 0 & 0\\
0 & 0 & 0
\end{array}\right]
\left[\begin{array}{c}
E_{x}+\imath B_{x} \\
E_{y}+\imath B_{y} \\
E_{z}+\imath B_{z} \\
\end{array}\right]
\end{eqnarray}

\noindent This transformation rotates both the $E$ and the $B$ field independently in the $X-Y$ plane. If we rotate along $3$ different axes then we get $3$ SU(3) generators as
\begin{eqnarray}
\left[\begin{array}{ccc}
0 & -i & 0\\
i & 0 & 0\\
0 & 0 & 0
\end{array}\right]\,,\qquad
\left[\begin{array}{ccc}
0 & 0 & 0\\
0 & 0 & -i\\
0 & i & 0
\end{array}\right]\,,\qquad
\left[\begin{array}{ccc}
0 & 0 & -i\\
0 & 0 & 0\\
i & 0 & 0
\end{array}\right]\,.
\end{eqnarray}

The third possibility is that while rotating $\vec{F}$ by an angle $\phi_z$ along $Z$ axis, instead of rotating $E_x \-- E_y$ and $B_x \-- B_y$, lets rotate between the coordinate $E_x \-- B_y$ and $E_y \-- B_x$. This type of transformation can also lead us to Eq.~\ref{Ftransformation}. 
This can be done by making the rotation matrix as

\begin{eqnarray}
\left[\begin{array}{c}
E'_{x}+\imath B'_{x} \\
E'_{y}+\imath B'_{y} \\
E'_{z}+\imath B'_{z} \\
\end{array}\right]
= 
\left[\begin{array}{ccc}
\cos(\phi_z) & i\sin(\phi_z) & 0\\
i\sin(\phi_z) & \cos(\phi_z) & 0\\
0 & 0 & 1
\end{array}\right]
\left[\begin{array}{c}
E_{x}+\imath B_{x} \\
E_{y}+\imath B_{y} \\
E_{z}+\imath B_{z} \\
\end{array}\right]\,.
\end{eqnarray}

\noindent It is easy to see that the total field strength $\vec{F}$ under such transformation rotates as expected. For small rotations, we can write the change as 
\begin{eqnarray}
\delta \left[\begin{array}{c}
E_{x}+\imath B_{x} \\
E_{y}+\imath B_{y} \\
E_{z}+\imath B_{z} \\
\end{array}\right]
= 
\imath \phi_z\left[\begin{array}{ccc}
0 & 1 & 0\\
1 & 0 & 0\\
0 & 0 & 0
\end{array}\right]
\left[\begin{array}{c}
E_{x}+\imath B_{x} \\
E_{y}+\imath B_{y} \\
E_{z}+\imath B_{z} \\
\end{array}\right]
\end{eqnarray}

\noindent Therefore, rotating along $3$ different axes give us the final $3$ generators of SU($3$), which are 
\begin{eqnarray}
\left[\begin{array}{ccc}
0 & 0 & 0\\
0 & 0 & 0\\
0 & 0 & 0
\end{array}\right]\,,\qquad
\left[\begin{array}{ccc}
0 & 0 & 1\\
0 & 0 & 0\\
1 & 0 & 0
\end{array}\right]\,,\qquad
\left[\begin{array}{ccc}
0 & 0 & 0\\
0 & 0 & 1\\
0 & 1 & 0
\end{array}\right]
\end{eqnarray}

\noindent This particular kind of transformation can be considered as the boost in classical electromagnetism.

Through this analysis, we try to explain the rotations in a complex field using an Electromagnetic field as an example to depict a visual picture of SU($3$) rotation. This can help the readers to visualize the rotation of the 1-forms and the 2-forms in the internal coordinate space.

\bibliographystyle{plainnat}
\bibliography{references}

\end{document}